\documentclass[JHEP,11pt]{article}
\usepackage{verbatim}

\usepackage{dsfont}
\usepackage{sidecap}
\sidecaptionvpos{figure}{t}
\usepackage{subfigure}
\usepackage[utf8]{inputenc}
\usepackage[english]{babel}
\usepackage{empheq}
\usepackage{amssymb,amsmath}
\usepackage{graphicx, xcolor, varwidth}
\usepackage{setspace}
\usepackage[permil]{overpic}
\usepackage{graphicx}
\usepackage{color}
\usepackage{braket}
\usepackage{simplewick}
\usepackage{jheppub}
\usepackage[T1]{fontenc}
\usepackage[autostyle]{csquotes}
\usepackage{ulem}
\usepackage{float}
\usepackage[bottom]{footmisc}

\usepackage{simpler-wick}

\usepackage{feynmp}
\colorlet{darkblue}{blue!70!black}
\colorlet{darkgreen}{green!70!black}

\numberwithin{equation}{section}

\numberwithin{equation}{section}

\newcommand{\mA}{\mathcal{A}}

\newcommand{\mO}{\mathcal{O}}

\newcommand{\mF}{\mathcal{F}}

\newcommand{\dl}{\delta \ell}
\newcommand{\tF}{\tilde{F}}



		\title{	{Scattering strings off quantum extremal surfaces }}

			\author[a]{Venkatesa Chandrasekaran}
			\author[b]{Thomas Faulkner}
			\author[c]{Adam Levine}

	\affiliation[a]{Center for Theoretical Physics and Department of Physics,
University of California, Berkeley, CA 94720, USA}
    \affiliation[b]{Department of Physics, University of Illinois, 1110 W. Green St., Urbana IL 61801-3080, U.S.A}
	
		\affiliation[c]{Institute for Advanced Study, Princeton, NJ 08540, USA}

		\emailAdd{arlevine@ias.edu}
		\emailAdd{tomf@illinois.edu}
		\emailAdd{ven\_chandrasekaran@berkeley.edu}
			
		\abstract{We consider a Hayden \& Preskill like setup for both maximally chaotic and sub-maximally chaotic quantum field theories. 
		We act on the vacuum with an operator in a Rindler like wedge $R$ and transfer a small subregion $I$ of $R$ to the other wedge. The chaotic scrambling dynamics of the QFT Rindler time evolution reveals the information in the other wedge. The holographic dual of this process involves a particle excitation falling into the bulk and crossing into the entanglement wedge of the complement to $r=R \backslash I$.
		With the goal of studying the locality of the emergent holographic theory we compute various quantum information measures on the boundary that tell us when the particle has entered this entanglement wedge. In a maximally chaotic theory, these measures indicate a sharp transition where the particle enters the wedge
		exactly when the insertion is null separated from the quantum extremal surface for $r$. For sub-maximally chaotic theories, we find a smoothed crossover at a delayed time given in terms of the smaller Lyapunov exponent and dependent on the time-smearing scale of the probe excitation.
		The information quantities that we consider include the full vacuum modular energy $R \backslash I$ as well as the fidelity between the state with the particle and the state without.
		Along the way, we find a new explicit formula for the modular Hamiltonian of two intervals in an arbitrary 1+1 dimensional CFT to leading order in the small cross ratio limit. We also give an explicit calculation of the Regge limit of the modular flowed chaos correlator and find examples which do not saturate the modular chaos bound. Finally, we discuss the extent to which our results reveal properties of the target of the probe excitation as a ``stringy quantum extremal surface'' or simply quantify the probe itself thus giving a new approach to studying the notion of longitudinal string spreading.
		
	    }

\addtocounter{page}{1}

\begin{document}
\maketitle 

\section{Introduction}\label{sec:intro}
Quantum extremal surfaces exist in a variety of previously unforseen places \cite{Penington:2019npb, Almheiri:2019vm, Engelhardt:2021un}. The position of these surfaces must obey constraints set by the scrambling of quantum information, as first detailed in the work of Hayden \& Preskill \cite{Hayden:2007wr}. For example, in the context of an AdS black hole evaporating into a bath, Hayden \& Preskill tell us that the quantum extremal surface associated to the post-Page time radiation must be null separated from a slice on the AdS boundary which is a scrambling time in the past. Since theories of gravity are expected to be dual to quantum systems which saturate the bound on chaos \cite{Maldacena:2015vu,Shenker:2013pqa}, a natural question is what happens to the bulk picture of a quantum extremal surface sitting a scrambling time into the past if the dual quantum system is \emph{sub-maximally} chaotic.\footnote{This question was first presented to the authors by Douglas Stanford and Ahmed Almheiri.} 
\begin{figure}
 \centering
 \includegraphics[width = .6\textwidth]{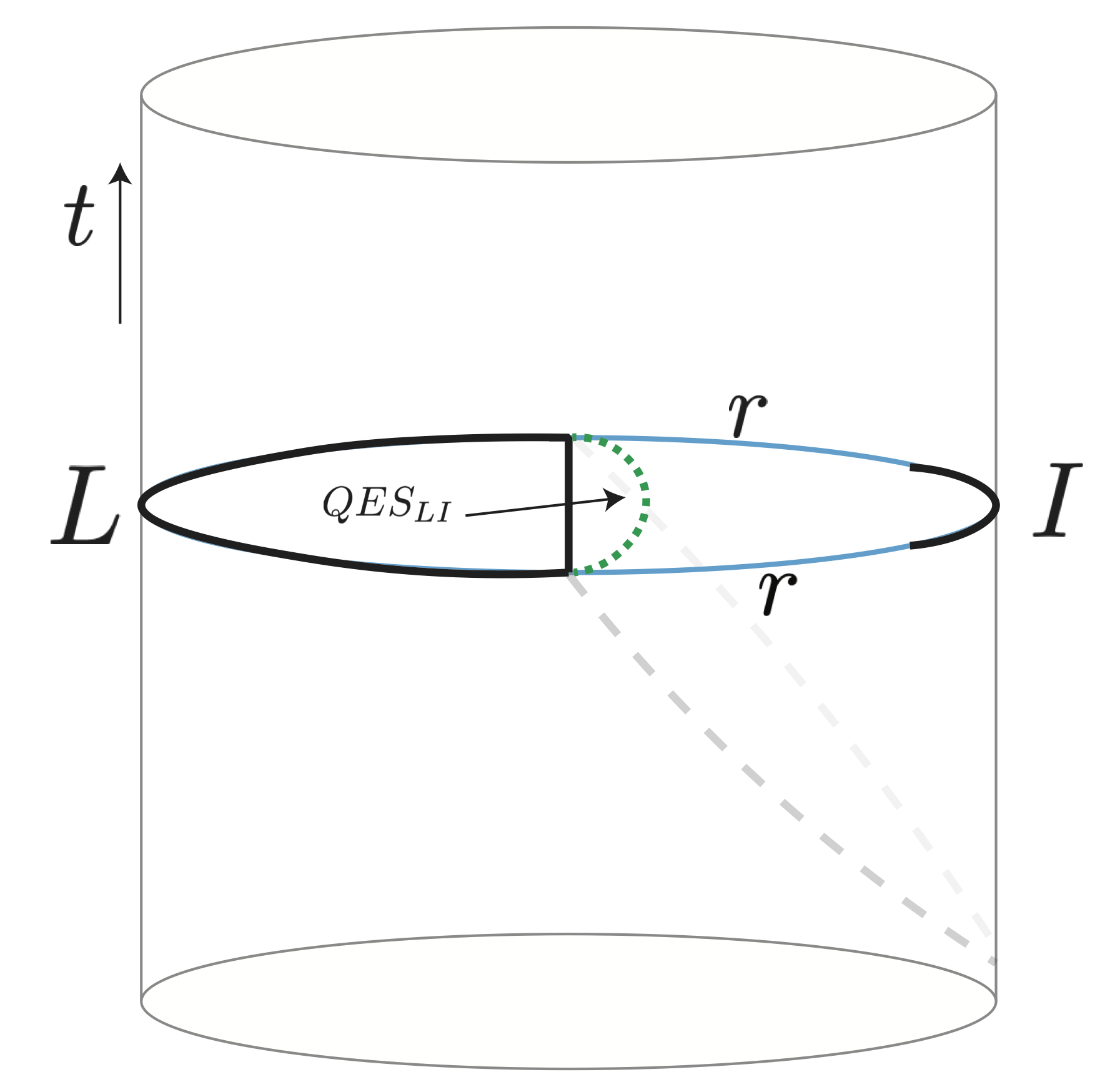}
 \caption{We consider a region $LI$ which is given by the union of two disjoint boundary spheres. We will be particularly interested in the limit where the radius of $I$, $\dl$, is small. The distance between $L$ and $I$ is controlled by $r_I$.}
 \label{fig:regions3d}
\end{figure}
Sub-maximally chaotic quantum systems are expected to be dual to a bulk theory with a non-zero string length. Our question can then effectively be rephrased as: how does a non-zero string length affect our conventional notion of a quantum extremal surface? In this work, we will try to examine this question by computing various boundary quantities that can sense the presence of an excitation in the entanglement wedge of some boundary region. This will allow us to probe the transition between when the particle enters or leaves the entanglement wedge.

To make our calculations concrete, we will focus on a scenario inspired by the work of Hayden \& Preskill \cite{Hayden:2007wr}. Consider a CFT in the vacuum state in any dimension. Take the region which is given by the union of a half space at $t=0$, $L = \lbrace (t=0,x<0,y^i)\rbrace$ and a small sphere $I$ of radius $\dl$ in the complementary half-space, $\bar{L} = R$. The sphere is centered about the point $x = r_I$ and also about the origin in the transverse coordinates. We denote the complement by $\overline{LI} = r$.

We will then perturb the vacuum by inserting an excitation located at $x^+ = t+x = e^{T_R}$. We create this excitation by acting with a conformal primary $\phi_R$ to get the state
\begin{align}\label{eqn:phistate}
\ket{\phi_{\delta}} = Z_{\delta} \,\phi(x^-=-e^{-i\delta},x^+=e^{i\delta}) \ket{\Omega},
\end{align} 
where $\ket{\Omega}$ is the vacuum state. Note that to make the state have finite energy, we have included a small amount of Euclidean time evolution $\delta$, which can be implemented by acting with an imaginary boost  $\sigma_R^{\delta/2\pi} \otimes \sigma_L^{-\delta/2\pi} \equiv \Delta_R^{\delta/2\pi}$, where $\sigma_{R,L}$ is the vacuum reduced density matrix for $R,L$. For a brief review of the full modular operator see Appendix \ref{app:tomitareview}.

\begin{figure}
 \centering
 \includegraphics[width = \textwidth]{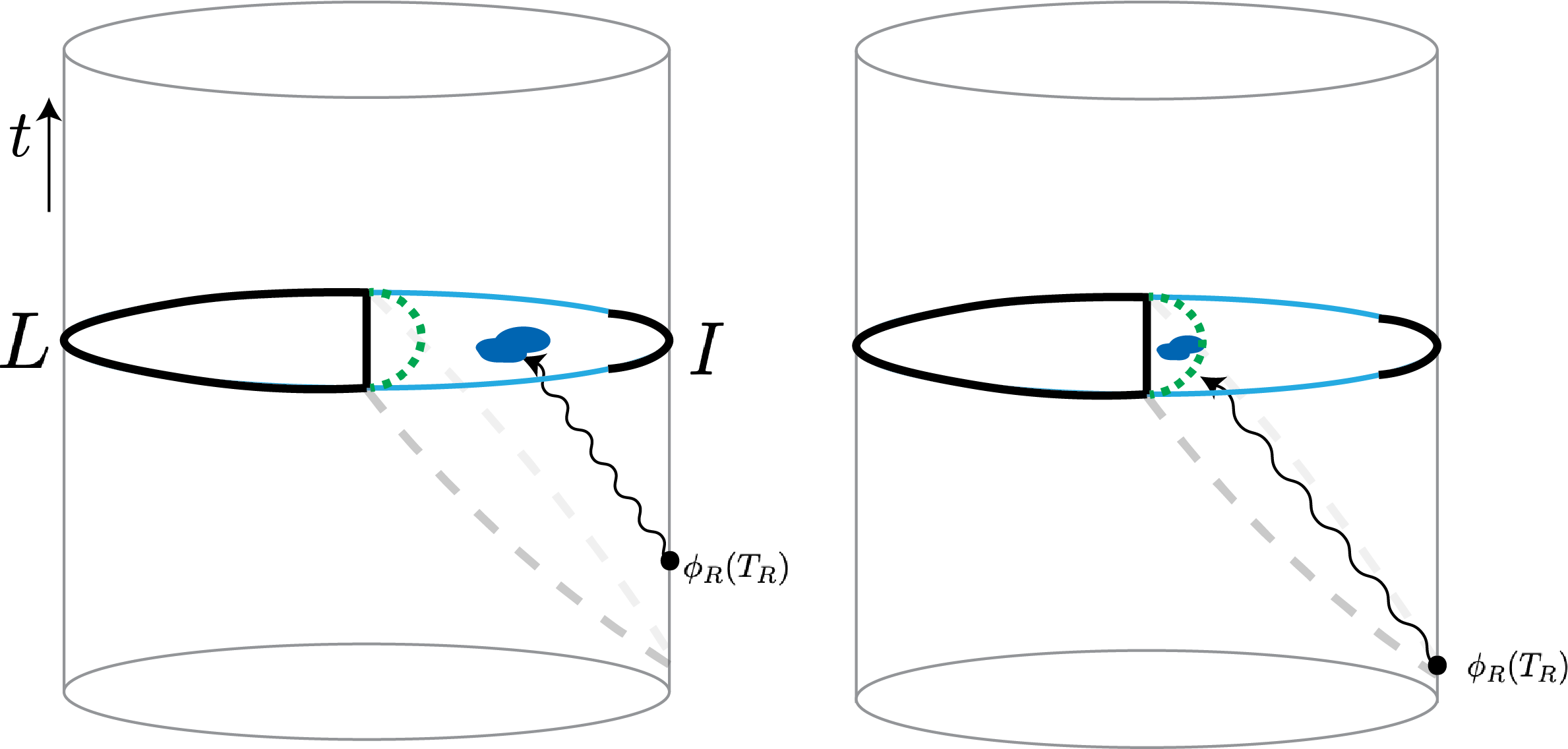}
 \caption{As we evolve the particle back in time, eventually it will pass from the entanglement wedge of $r$ to the entanglement wedge of $LI$.  \label{fig:3dtransition}}
\end{figure}

We then evolve this state back in time using the boost generator
\begin{align}
\ket{\phi_{\delta}} \to \Delta_R^{iT_R/2\pi} \ket{\phi_{\delta}}.
\end{align}
At some point this excitation will move into the past of the sphere $I$. By boost symmetry of the vacuum, the excitation in the bulk can always be localized within the quantum extremal wedge of the boundary region $R$. On the other hand, the quantum extremal wedge of $r$ is actually strictly smaller than that of $R$. Indeed, the excitation will only remain in the quantum extremal wedge of $r$ until some (large) value of $T_R$. If we assume that the boundary operator is dual to a massless field in the bulk, eventually the excitation will pass cleanly from the wedge of $r$ to the quantum extremal wedge of $LI$, as illustrated in Figure \ref{fig:3dtransition}. While this setup can be analyzed in any dimension, for simplicity we will work with a toy model in two dimensions first introduced by \cite{AMM}. The model involves two dimensional JT gravity coupled to a 1+1 dimensional CFT. The gravitational theory is coupled to a non-gravitational bath at finite temperature. In this setup, the region $I$ is taken to be an interval in the right bath while the region $L$ is taken to be the whole left bath plus the left black hole exterior (see Figure \ref{fig:ads2regions}). In Section \ref{sec:discussion}, we will come back to higher dimensions and argue that our results are qualitatively unchanged.

Consider the fidelity on the region $r$ between the state
\begin{align}\label{eqn:psistate}
    \ket{\psi_{\lambda}} =  \cos(\lambda) \ket{\Omega} + i \sin(\lambda) \ket{\phi_\delta}
\end{align} 
and the vacuum $\ket{\Omega}$,
\begin{equation}\label{eqn:fidelityintro2}
F(\psi_{\lambda}|\Omega; r)
\equiv  \sup_{U_{LI}} \left| \Braket{\psi_{\lambda}| U_{LI}|\Omega} \right|^2,
\end{equation}
where $U_{LI}$ is an arbitrary unitary on $LI$.
For fixed non-zero $0<\lambda<\pi/2$ the fidelity in \eqref{eqn:fidelityintro2} tells us the distinguishability of these two states reduced to $r$. If the fidelity is $1$ then we can safely conclude that the excitation has left the $r$ entanglement wedge. 

It is much simpler to compute the small $\lambda$ limit of this fidelity. 
In this limit, the fidelity is close to one. Namely, 
\begin{align}
F(\psi_{\lambda}|\Omega; r) = 1 - \lambda^2 \chi(\psi,\Omega;r) + \mathcal{O}(\lambda^3)
\end{align} where $\chi$ is often referred to as the \emph{fidelity susceptibility}. If $\chi$ vanishes, then we again conclude that the excitation has left the $r$ entanglement wedge.\footnote{This conclusion is less obvious than at finite $\lambda$, see Section \ref{sec:2}.}
As we will see explicitly in the 2-d toy model first discussed in \cite{AMM}, for a maximally chaotic boundary theory the fidelity susceptibility transitions to zero when the boundary operator $\phi_R(T_R)$ becomes null separated from the quantum extremal surface of $LI$ in the bulk.\footnote{After submission of this work, the authors were made aware of previous results using the fidelity of two holographic CFT states to image extremal surfaces in the bulk \cite{Kusuki:2019aa, Suzuki:2019aa}. While these two works focused on the classical position of the extremal surface, we concern ourselves here with the $1/N$ or quantum corrections to the quantum extremal surface as well as the finite coupling corrections to the fidelity. We thank Tadashi Takaynagi for discussions on this point.}

We also compute the susceptibility for boundary theories which are dual to bulk theories with a non-zero string length.\footnote{Strictly speaking, we will use a phenomenological model put forth in \cite{MSY-2} for the effect of non-zero string length. We discuss more realistic models briefly in Section \ref{sec:discussion}.} For a theory with sub-maximal chaos which is expected to be dual to a bulk theory with non-zero string length, we find several differences with the gravitational case. In particular, the transition between the excitation being in the entanglement wedge of $r$ and being in the entanglement wedge of $LI$ is less sharp than for a theory with zero string length. Moreover, in the sub-maximally chaotic case the transition time itself is shifted by a $\delta$-dependent amount. In \ref{sec:bulkinterp}, we will provide a speculative interpretation of these results in terms of stringy physics in the bulk. We now provide a detailed overview of the main points of the paper which should be sufficient for the reader uninterested in computational details.

\subsection*{A detailed summary of results}

A cruder but more easily computable diagnostic of when the probe excitation has left the entanglement wedge of $r$ is the difference
\begin{align}\label{eqn:sdiff}
S( \psi_{\lambda} | \Omega; r) - S(\psi_{\lambda} |\Omega; LI) = -\sin^2(\lambda)\braket{\log \Delta_r}_{\phi_{\delta}}
\end{align}
where $S(\phi | \Omega; B)$ is the relative entropy between the states $\ket{\psi_{\lambda}}$ defined in \eqref{eqn:psistate} and the vacuum $\ket{\Omega}$ reduced to the region $B$. Here the operator $\log \Delta_r = -\log \sigma_r +\log \sigma_{LI}$ is the vacuum modular operator for the region $r$. For regions which are Rindler wedges, $\log \Delta_r$ generates boosts. For more general regions, $\log \Delta_r$ generates complicated non-local flows \cite{Arias:2018aa}. Note that in the equality in \eqref{eqn:sdiff}, we have used purity of the global state together with the definition of the relative entropy as
\begin{align}
S(\phi |\Omega; r) = \text{Tr}[\rho_r \log \rho_r] - \text{Tr}[\rho_r \log \sigma_r].
\end{align} 
where $\rho_X$ is the reduced density matrix for subregion $X$ and global state $\ket{\phi}$.
We will refer to the quantity in \eqref{eqn:sdiff} as the \emph{full modular energy}. The relative entropy is a good measure of the distinguishability of two quantum states; the relative entropy is non-negative and zero if and only if $\rho_r = \sigma_r$. 
Furthermore, when the relative entropy is large, the states are in some sense easier to distinguish operationally. Such boundary states accordingly have ``more'' differing states of the bulk quantum fields on their entanglement wedges via the equivalence between bulk and boundary relative entropies \cite{Jafferis:2015del}. Thus, one might expect that when the difference in \eqref{eqn:sdiff} is positive, the excitation is mostly contained in the entanglement wedge of $r$ and when it is negative the excitation is mostly contained in the entanglement wedge of $LI$. The quantity in \eqref{eqn:sdiff} has the advantage that it is simpler to compute. The downside is that there is no rigorous reason (that we know of) for why such a quantity should encode whether an excitation is in the entanglement wedge or not.

In order to compute this quantity, we will need the form of the \emph{half-sided modular Hamiltonian} for the region $LI$, $H_{LI}$, where $-\log \Delta_r = 2\pi (H_r - H_{LI})$. We thus begin in \textbf{Section \ref{sec:3}} by computing $H_{LI}$ via a replica trick in the limit where the size of $I$, $\dl$, is small.  The modular Hamiltonian for $LI$ can be computed via a replica trick where twist operators are inserted at the boundaries of $LI$. Since $I$ is small, we can do a twist OPE and consider the leading contribution to the OPE, which will be governed by the lightest operators, $O$, in the spectrum of the CFT, as was the case in the work of \cite{Agon:2015ftl}. Roughly speaking, what we find is that $H_{LI}$ is quadratic in this lightest operator
\begin{align}\label{eqn:modham}
H_{LI} = H_L^{loc} + H_I^{loc} + \int_{LI} dx dy\, f(x,y) \,O(x) O(y) + \mO((\dl)^{2\Delta_O +1})
\end{align}
where $f(x,y) \sim (\dl)^{2\Delta_O}$ with $\Delta_O$ the conformal dimension of $O$. $H_{L,I}^{loc}$ are the vacuum modular Hamiltonians for $L,I$ respectively.  In a CFT, these Hamiltonians generate a local flow and so are just given by integrals of the stress tensor over the region. In Section \ref{sec:3}, we will find explicit forms for $f(x,y)$ in the small $\dl$ limit. It is interesting to note the similarity between the Hamiltonian discussed in \eqref{eqn:modham} and those discussed in the context of traversable wormholes \cite{GJW,Maldacena:2018lmt}, where a bilocal interaction was added between the two sides of the thermo-field double. Here the analog of the thermo-field double is just the vacuum, viewed as thermally entangled between two Rindler wedges. In the bulk dual of maximally chaotic theories, the action of unitaries generated by bilocal Hamiltonians such as in \eqref{eqn:modham} is often just a simple null shift as discussed in \cite{LMZ}. We will find this to be the case in our work as well. The result \eqref{eqn:modham} is also reminiscent of the connection between modular Hamiltonians and particle worldline time evolution as discussed in \cite{Jafferis:2020ora}. 

With equation \eqref{eqn:modham} in hand, we consider a toy model in \textbf{Section \ref{sec:4}} which will allow us to compute the full modular energy in \eqref{eqn:sdiff}. The setup will be to consider a theory dual to a 1+1 dimensional CFT coupled to JT gravity on $\text{AdS}_2$. The gravitating region will also be coupled to a flat space bath at finite temperature. This setup was first considered in \cite{AMM, AMMZ}. In this case, the expectation value of the full modular operator in \eqref{eqn:sdiff} is given by an out-of-time-order correlator (OTOC) of the type studied in \cite{MSY-2,Shenker:2013pqa}. We will find that the answer is (up to overall coefficients)
\begin{align}\label{eqn:gravsdiff}
S(\psi_{\lambda}|\Omega; r) - S(\psi_{\lambda} | \Omega; LI) \sim \frac{\Delta_{\phi}\sin^2(\lambda)}{\delta} \left(1-e^{-T_R}\delta x_Q^+\right),
\end{align}
where $\delta x_Q^+$ is the null shift, in Kruskal coordinates, of the quantum extremal surface for region $LI$ due to the inclusion of $I$. Here $\Delta_{\phi}$ is the dimension of the probe operator $\phi$ and $\delta$ is the amount of Euclidean evolution introduced in \eqref{eqn:phistate}. We see immediately from this formula that the difference in relative entropies changes sign precisely when the excitation is null separated from the vacuum entanglement wedge for $r$ (which is the same as the entanglement wedge for $LI$).

To get this formula, we had to assume the boundary theory was maximally chaotic. We can mock up the case where the boundary theory has a sub-maximal Lyapunov exponent by following \cite{MSY-2}. The OTOC needed to compute \eqref{eqn:sdiff} can be calculated via a bulk scattering amplitude in the eikonal limit. The gravitational scattering matrix in the eikonal limit is just a simple phase in momentum space, namely $e^{i\delta(s)} = e^{-iG_Np_+ q_-}$ where $p_+$ and $q_-$ are the null momenta of the particles being scattered \cite{Shenker:2014tu,HOOFT198761,Verlinde_1992,Kabat:1992tb}. We will model the stringy scattering amplitude by modifying the scattering phase to have the form $e^{i\delta_J} \equiv e^{-G_N(ip_+q_-)^{J-1}}$, where $1\leq J\leq 2$ is the so-called Pomeron spin. A theory with maximal chaos has $J = 2$, which is the spin of a graviton. On the boundary side, we can think in terms of a conformal diagram between the $\psi$ and $O$ operators. In the maximally chaotic case, the leading contribution to the diagram comes from stress tensor exchange. In the sub-maximally chaotic case, the exchange includes an effective ``Pomeron'' operator which resums the lightest Regge trajectory. The stringy physics is encoded in this Pomeron exchange \cite{Chew:1961ev, Gribov:1961ex, Brower:2007uc}. Using this modified scattering phase, we find in Section \ref{sec:4}
\begin{align}\label{eqn:stringysdiff}
S(\psi_{\lambda}|\Omega; r) - S(\psi_{\lambda} | \Omega; LI) \sim \frac{\Delta_{\phi}\sin^2(\lambda)}{\delta} \left(1-\delta^{2-J}e^{-(J-1)T_R}\delta x(J,\Delta_{\phi})/2\right).
\end{align}
This formula has two interesting features:
\begin{itemize}
\item The change in sign occurs at an earlier time due to the smaller Lyapunov exponent. 
\item The exact time at which this transition occurs depends both on the smearing length $\delta$ and the dimension of the perturbing operator $\Delta_{\phi}$. For $J=2$, $\delta x(J=2)$ is independent of $\Delta_{\phi}$.
\end{itemize}

Having computed the full modular energy in \eqref{eqn:sdiff}, we turn to computing the fidelity susceptibility between the state $\ket{\psi_{\lambda}}$ defined in \eqref{eqn:psistate}, and the vacuum. We use the results of Hijano \& May \cite{May:2018ti} to calculate the fidelity (susceptibility) explicitly. In order to compute the susceptibility, we find that we need to first compute the modular flowed correlation function
\begin{align}\label{eqn:correlator}
\mathcal{F}(t_L, t_R;s)  = \langle \rho^{is/2\pi}_{LI}\rho^{-is/2\pi}_L \phi_L(t_L)\rho^{is/2\pi}_L \rho^{-is/2\pi}_{LI}\phi_R(t_R)\rangle 
\end{align}
in the limit of large $t_L-t_R$, which we will refer to as the \emph{Regge limit}. More explicitly, one can directly relate the susceptibility to an integral over modular flow parameter, $s$, of the correlator in \eqref{eqn:correlator}. See Appendix \ref{app:susceptibility} for details. Similar correlation functions to that in \eqref{eqn:correlator} were computed in the proof of the quantum null energy condition \cite{Balakrishnan:2017aa,Faulkner:2018vl}. We compute this correlation function in \textbf{Section \ref{sec:5}} for theories with both maximal and sub-maximal chaos. It was shown in \cite{Faulkner:2018vl, Boer:2019td} that correlation functions like that in \eqref{eqn:correlator} obey a version of the chaos bound, where the modular evolution parameter $s$ plays the role of time. We find that for the correlator in \eqref{eqn:correlator}, the modular Lyapunov exponent is just the same as the more standard Lyapunov exponent defined in terms of out-of-time-order correlators of local operators. Furthermore, we argue that $\mathcal{F}$ must obey a reality condition in the complex $s$-plane and we prove that our answer obeys this reality condition for $1\leq J \leq 2$.

In \textbf{Section \ref{sec:6}}, we use the results for the correlator in \eqref{eqn:correlator} to compute the fidelity susceptibility $\chi(\psi_{\lambda},\Omega)$. We argue that for probe operators of integer dimension $\Delta_{\phi}$, as $J \rightarrow 2$ the fidelity approaches a step function in the limit where the smearing scale $\delta$ goes to zero. For the sub-maximally chaotic case, we compute the fidelity susceptibility when $\Delta_{\phi}=1/2$ in the small $\delta$ limit and at large $e^{-T_R}$. We find that the fidelity takes the form
\begin{align}\label{eqn:stringyfidelity}
    F(\psi_{\lambda} |\Omega; r) \sim 1-2\lambda^2 \left( \exp\left(- (c \ \delta x(J) \delta^{2-J}e^{-T_R(J-1)})^{\frac{1}{2-J}}\right) + \mathcal{O}(\delta)\right) + \mathcal{O}(\lambda^3)
\end{align}
where $c$ is an order unity numerical coefficient and $\delta x(J) = \delta x(J,\Delta_{\phi} = 1/2)$ is the same number that appears in \eqref{eqn:stringysdiff}. We also can compute the fidelity numerically for special $J$, namely $J = 3/2$. We find numerical agreement with \eqref{eqn:stringyfidelity}. We see that the fidelity approaches the value one more slowly and the time at which it gets close to one depends logarithmically on $\delta$ at small $\delta$. This agrees with the dependence on $\delta$ of the turnover time, $T_R^*$, where \eqref{eqn:stringysdiff} becomes negative. Furthermore, we see that the turnover gets sharper as $J$ approaches $2$, which is the gravity limit. Our numerical results are presented in Figures \ref{fig:mainresultstringy} and \ref{fig:mainresultgrav}. 

In \textbf{Section \ref{sec:bulkinterp}}, we discuss possible bulk interpretations of these results. We firstly discuss how we might interpret our results in terms of a ``stringy'' notion of the quantum extremal surface. Since the bulk theory no longer has a sharp notion of locality we expect a stringy QES to also not be local - and indeed this is indicated by the smoothed transition of $\chi$ as a function $T_R$. However we also point out that such an interpretation cannot completely account for our computation - in particular we cannot account for the strong dependence of the crossover time on the smearing scale $\delta$.
As a counterpoint, we then discuss whether our results could instead be interpreted in terms of longitudinal string spreading, first predicted in \cite{Susskind:1994vn} and explored in \cite{Dodelson:2015un,Dodelson:2017ub,Mousatov:2020wi,Dodelson:2017wm}. Our results suggest that in a theory with non-zero $\ell_s$ the probe string remains in the entanglement wedge of $r$ for a longer time. This might suggest that the string's wavefunction is more spread out in the longitudinal direction (the $x^+$ direction in Kruskal coordinates). One can try to read off the longitudinal size of the excitation by finding a region $LI$ from which the excitation is just barely reconstructable. A naive estimate on the size of the string is then $\delta x_{\text{string}}^+ \sim \delta x^+_Q(LI)-e^{T_R}$. We find 
\begin{align}
    \delta x^+_{\text{string}} \sim (2-J)e^{T_R}\log \left(\frac{1}{r_I \delta e^{T_R}}\right) + \mathcal{O}((2-J)^2).
\end{align} 
Remembering that $2-J$ is proportional to the string length in more realistic models, we see that this formula describes an object which is spreading logarithmically but also length contracting. In this case the length contraction wins out. We compare this with previously discussed calculations of longitudinal string spreading in \cite{Susskind_1993,Larsen_1999}.

Finally in \textbf{Section \ref{sec:discussion}}, we end with a discussion of how our results generalize to more realistic models in higher dimensions. We also comment on a connection between the modular flowed correlator in \eqref{eqn:correlator} and the expectation value of continuous spin null energy operators in modular flowed states.


\section{The Setup}\label{sec:2}

\begin{figure}
 \centering
 \includegraphics[width = .5\textwidth]{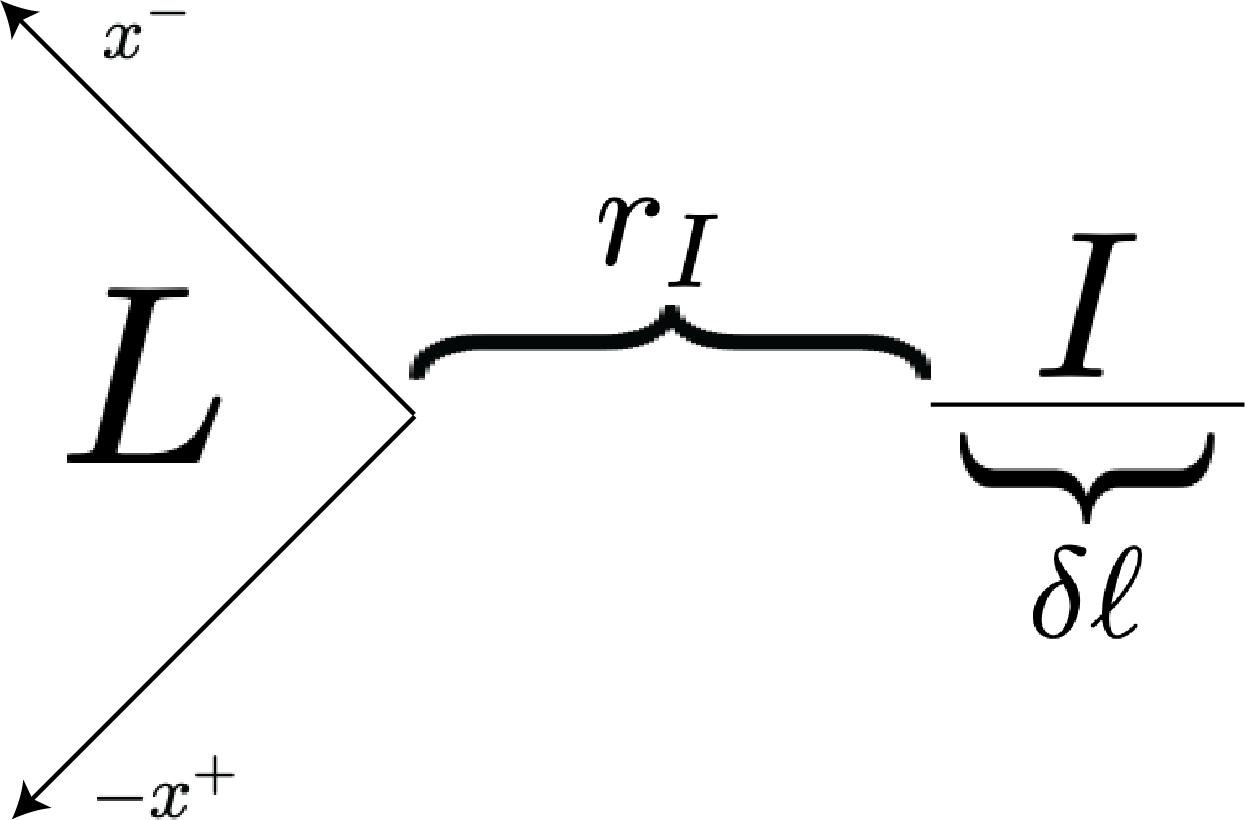}
 \caption{The setup we consider in a 1+1 dimensional CFT.}
 \label{fig:regions3dparticle}
\end{figure}

In this section we describe the setup which will be used throughout this work. We also fix the notation and conventions adapted to this setup. Our goal in this work is to compute a boundary quantity which is sensitive to the position of a quantum extremal surface in the bulk. To accomplish this goal, we will consider a variant of the Hayden \& Preskill \cite{Hayden:2007wr} protocol, adapted to the vacuum state of a CFT in arbitrary dimensions, which we now describe.

Consider the vacuum state of a ${\rm CFT}_d$ on $\mathbb{R}^{d-1,1}$. We will be interested in studying the quantum extremal surface in the bulk associated to the boundary region, which we will denote $LI \equiv L \cup I$, where $L$ is the Rindler wedge $L = \lbrace{(t,x,y^i) |\, x\leq 0,\  x \leq t \leq -x\rbrace}$ with $i = 1,... d-2$. The region $I$ will be the domain of dependence associated to a sphere at $t =0$ of radius $\dl/2$ centered at $y^i = 0$ and $x = r_I$. We will denote the complementary region of $LI$ as $\overline{LI} =r $. We will also denote the complement of $L$ by $R$. The notation is laid out in Figure \ref{fig:regions3dparticle}.

In this paper, we consider the ``long distance'' limit, where $\dl/r_I \ll 1$, as a simplifying limit. We expect the qualitative features of our results to generalize away from this limit. For small $\dl/r$, the quantum extremal surface (QES) associated to $LI$ will be slightly displaced from the AdS-Rindler horizon. The displacement will be, to leading order, determined by the entanglement due to bulk quantum fields. In principle, the position of the QES can be computed by solving the equation
\begin{align}
\frac{\theta^{(+)}}{4G_N} = - \frac{1}{\sqrt{h}} \frac{\delta S_{bulk}}{\delta X^+}
\end{align}
where $\theta^{(+)}$ is the null-expansion of QES. Here, the position of the surface can be described via embedding functions $X^{\pm}(\sigma)$ with $\sigma$ the internal coordinates of the surface and $h$ is the induced area element of the surface. Entanglement wedge nesting \cite{maximin} dictates that the QES of LI must be as large as that of $L$ alone. Since this statement is only saturated in highly symmetric cases, we expect the QES of $LI$ to be strictly larger than that of $L$, which we shall see explicitly.

Modeling after the Hayden \& Preskill protocol, we drop a message (particle) from the far past into the bulk. To determine when the message is reconstructable from $LI$ we consider a code subspace which is spanned by the vacuum and the state $\ket{\phi_{\delta}}$ defined in \eqref{eqn:phistate}, which we repeat here
\begin{align}
&\ket{\phi_{\delta}} = Z_{\delta} \Delta_{R}^{\delta}\phi_R(T_R)\ket{\Omega}.
\end{align}
Here $\ket{\Omega}$ is the vacuum state, $\phi_R$ is some conformal primary of dimension $\Delta_{\phi}$ and $Z_{\delta} = (2\sin(\delta))^{2\Delta_{\phi}}$ is the normalization. We assume that $\Delta_{\phi}$ is small enough so that we can neglect backreaction due to this insertion. 

Note that we are evolving $\phi$ with the boost generator around $x^- = x^+=0$, which is $K_R$. We have included some amount of Euclidean time evolution to make $\ket{\psi}$ normalizable. For small $\delta$, one can replace this Euclidean evolution with a small amount of Lorentzian time smearing to project out high energy contributions. 

At $T_R=0$, $\phi_R(T_R)$ is an operator in\footnote{This operator is unbounded so it does not belong to the usual notion of an algebra, mathematically it should be thought of as being affiliated to the algebra - meaning that the spectral projections are in the algebra. We will not worry about making this distinction moving forward. } the commutant of $LI$, assuming $0<X_R< r-\dl/2$. In other words, $\phi_R(T_R = 0)$ is in the algebra associated to $r= \overline{LI}$. When $T_R$ is large and negative, we can imagine that this insertion corresponds to a shockwave in the bulk propagating along a light ray. Such states have been discussed before in \cite{Afkhami-Jeddi:2017aa}. As we move $T_R$ to more and more negative values, eventually the shockwave will cross out of the entanglement wedge of $r$. Our goal is to devise a boundary quantity which senses when this transition occurs. Our main tool will be entanglement wedge reconstruction \cite{Faulkner:2017vdd} interpreted in an error correcting language \cite{Cotler:2017erl}.

\subsection{Reconstructing the excitation}

To identify when the particle leaves the entanglement wedge of $r$, we need to know when the two dimensional code subspace spanned by the vacuum and the state with the particle, $\ket{\phi_{\delta}}$, is reconstructable from $LI$. Note that by symmetry, $\braket{\phi_{\delta}|\Omega}=0$. To quantify when this qubit's worth of information is reconstructable from $LI$, we can imagine maximally entangling a qubit in Eve's system with the qubit in the $LR$ system to form the state
\begin{equation}
\left| \Phi_{LIrE}(T_R)\right> = \frac{1}{\sqrt{2}} \left( \left| 0 \right>_{E} \left| \Omega \right>_{LIr} +\left| 1 \right>_{E} \left| \phi_{\delta}(T_R) \right>_{LIr} \right).
\end{equation}
If the mutual information $I(E, r)_\Phi \approx 0$ then there exists an isometry $V_{LI \rightarrow IL \widetilde{E} \widetilde{E}'}$ that approximately extracts the entanglement with the references on the $LI$ Hilbert space:
\begin{equation}
\label{eq:VLI}
(V_{LI \rightarrow LI \widetilde{E} \widetilde{E}'})
\left| \Phi_{LIrE} \right>
\approx \left| E \widetilde{E}' \right>
\otimes \left| \Phi_{\widetilde{E} r IL} \right>
\end{equation}
with $\left| E \widetilde{E}' \right> =\frac{1}{\sqrt{2}}\left( \left| 0 0 \right> + \left| 11 \right> \right)$. The existence of such an isometry guarantees that any single qubit operator on Eve's system can be represented as an operator acting only on $LI$. Unfortunately, computing $I(E,r)_{\Phi}$ as a function of $T_R$ is difficult so in this work we opt for studying a different object.

To this aim, consider the converse: if we know that the logical $X$ and $Y$ operators, defined as
\begin{align}\label{eqn:pauli}
&\sigma_X = \ket{\phi_{\delta}}\bra{\Omega} + \ket{\Omega}\bra{\phi_{\delta}}\nonumber \\
&\sigma_Y = i\left(\ket{\phi_{\delta}}\bra{\Omega} - \ket{\Omega}\bra{\phi_{\delta}}\right),
\end{align} can be reconstructed on the code-subspace from $LI$, then we can conclude that $I(E:r)_{\Phi} = 0$ \cite{Harlow:2016vwg}. To find when $\sigma_X$ and $\sigma_Y$ are reconstructable on $LI$, consider the fidelity on the region $r$ between the states
\begin{align}\label{eqn:psistatenew}
    \ket{\psi^{X,Y}_{\lambda}} = e^{i\lambda \sigma_{X,Y}} \ket{\Omega}
\end{align} 
and the vacuum $\ket{\Omega}$,
\begin{equation}
\label{eqn:fidelity}
F(\psi_{\lambda}^{X,Y}|\Omega; r)
\equiv  \sup_{U_{LI}} \left| \Braket{\psi_{\lambda}^{X,Y}| U_{LI}|\Omega} \right|^2,
\end{equation}
where the previously defined $\psi_\lambda$ in \eqref{eqn:psistate} corresponds to $\psi_\lambda^X$.
We will often suppress the superscript on $\psi_{\lambda}$ when a given equation holds for both $\ket{\psi_{\lambda}^X}$ and $\ket{\psi_{\lambda}^Y}$.
The fidelity in \eqref{eqn:fidelity} tells us how well we can reconstruct the action of $\sigma_{X,Y}$ when just given access to unitaries on $LI$. 

In more detail, when the bulk excitation is well localized within the QES of $LI$, it should be reconstructable from $LI$. This means that for both logical unitaries $e^{i \lambda \sigma_{X,Y}}$, where $\sigma_X$ and $\sigma_Y$ are the two Pauli operators on the two-dimensional code subspace spanned by $\ket{\Omega}$ and $\ket{\phi_{\delta}}$, there should exist unitaries $U^{X,Y}_{LI}$ which act identically to $e^{i \lambda \sigma_{X,Y}}$ on the code subspace. In equations,
\begin{align}
\exists \ U^{X,Y}_{LI}\mid ~ U^{X,Y}_{LI} \ket{\Omega} \approx e^{i\lambda \sigma_{X,Y}}\ket{\Omega} \equiv \ket{\psi_{\lambda}^{X,Y}}.
\end{align}
It is clear from the definition in \eqref{eqn:fidelity} that this condition is then directly measured by the quantum fidelity between $\ket{\psi_{\lambda}}$ and $\ket{\Omega}$.

We can also consider taking the small $\lambda$ limit so that $\ket{\psi_{\lambda}^{X}} =\ket{\Omega} +i\lambda \ket{\phi_{\delta}} + \mathcal{O}(\lambda^2)$ and $\ket{\psi_{\lambda}^{Y}} =\ket{\Omega} -\lambda \ket{\phi_{\delta}} + \mathcal{O}(\lambda^2)$. In this limit, the fidelity is close to one. Namely, 
\begin{align}
F(\psi^{X,Y}_{\lambda}|\Omega; r) = 1 - \lambda^2 \chi(\psi^{X,Y},\Omega;r) + \mathcal{O}(\lambda^3)
\end{align} where $\chi$ is often referred to as the \emph{fidelity susceptibility}.\footnote{Note that the fidelity is always less than or equal to one and so only even powers of $\lambda$ can appear as the leading order correction to one.} When the fidelity susceptibility vanishes so that the fidelity is even closer to one, $F(\psi_{\delta}|\Omega) \approx 1- \mathcal{O}(\lambda^4)$, then we know that there exists a unitary $U_{LI}$ such that
\begin{align}\label{eqn:reconequality}
    U^{X,Y}_{LI}\ket{\psi^{X,Y}_{\lambda}} = \ket{\Omega} + \mathcal{O}(\lambda^2).
\end{align}
Since $\lambda$ is small, we can assume that the maximizing unitary $U^{X,Y}_{LI}$ can also be expanded in $\lambda$ as $U^{X,Y}_{LI} = 1+i\lambda \delta h_{X,Y} - \frac{\lambda^2}{2} (\delta h_{X,Y})^2+...$ with $\delta h_{X,Y}$ Hermitian. Plugging this into \eqref{eqn:reconequality} tells us that $\delta h_{X,Y} \ket{\Omega} = \sigma_{X,Y}\ket{\Omega}$. Strictly speaking, this equality is not enough to claim that we can reconstruct the whole two-dimensional code subspace. For that, we also need the equality $\delta h_{X,Y}^2 \ket{\Omega} = \ket{\Omega}$ so that we also get the algebra correct. To get this, one needs to argue that the fidelity is one up to order $\lambda^6$ corrections.\footnote{We leave as an exercise for the reader that this is sufficient to get the equality $\delta h_{X,Y}^2 \ket{\Omega} = \ket{\Omega}$.} Finding when $F = 1- \mathcal{O}(\lambda^6)$ requires computing the fidelity up to order $\lambda^4$ which is more involved.

In this work, we opt for just computing the fidelity susceptibility $\chi(\psi^{X,Y},\Omega)$ and looking for when it vanishes. As we will see, for a maximally chaotic boundary theory, we find that the fidelity susceptibility transitions to zero when the boundary operator $\phi_R(T_R)$ becomes null separated from the quantum extremal surface of $LI$ in the bulk. In the $\delta \to 0$ limit, the transition becomes sharp. Since we know from general reasoning about entanglement wedge reconstruction in gravitational theories that the order $\lambda^4$ term in the fidelity should also go to zero at this time, this suggests to us that the order $\lambda^4$ term in the fidelity is controlled by similar correlators as the susceptibility and so will vanish at the same time as the susceptibility regardless of the bulk string length.  In the remainder of this work, we will assume that the timescale for the transition of the order $\lambda^4$ term is the same as for the susceptibility. The reader should keep the caveat discussed here in mind. We could in principle prove this assumption by going to higher orders in $\lambda$ but we leave that for future work.

\subsection{Other information measures of the transition}

As discussed in the introduction, we will be interested in states which are perturbatively close to the vacuum $\lambda \ll 1$, and so we can use the work of \cite{May:2018ti} to expand the fidelity to leading order in $\lambda$. This will be the content of Section \ref{sec:6}. Since this is technically involved, however, we first introduce a different quantity which captures the same qualitative aspects of the transition between the particle existing in $EW(LI)$ versus $EW(r)$.

Consider the relative entropy defined as
\begin{align}
S(\rho| \sigma) = \text{Tr}[\rho \log \rho] - \text{Tr}[\rho \log \sigma].
\end{align}
The relative entropy $S(\rho| \sigma)$ is yet another measure of distinguishability between two states $\rho$ and $\sigma$. Indeed, the relative entropy bounds the fidelity \cite{Nielsen_2009}
\begin{align}\label{eqn:fidboundsrel}
1-F(\rho,\sigma) \leq e^{-\frac{1}{2} S(\rho | \sigma)}.
\end{align}
When $S(\rho| \sigma)$ is large, the two states are more easily distinguishable.\footnote{This statement should be treated with some caution: the relative entropy is sometimes a bit too fine grained a measure of distinguishability. The relative entropy can in principle be large even when the fidelity is close to one. The real operational distinguishability of the two states is given by the trace-norm distance, which upper and lower bounds the fidelity via the Fuchs Van de Graaf inequalities.} On the other hand, when $S(\rho|\sigma)$ is small, the states are indistinguishable. 

As described in the introduction, when $S(\rho^{\psi_{\lambda}}_r |\sigma^{\Omega}_r)$ is large, the excitation is well localized in $r$. Here $\rho^{\psi_{\lambda}}_r$ is the reduced density matrix of the state $\ket{\psi_{\lambda}}$ on $r$ and $\sigma_{\Omega}^r$ is the reduced density matrix of the vacuum on $r$. When the relative entropy on $r$ is small, we know from equation \eqref{eqn:fidboundsrel} that the excitation has left $EW(r)$. Thus one might expect that the quantity
\begin{align}
S(\rho^{\psi_{\lambda}}_r|\sigma_r)- S(\rho^{\psi_{\lambda}}_{LI}|\sigma_{LI})
\end{align}
is a measure of when the excitation has left the entanglement wedge of $r$. The point at which this quantity crosses from positive to negative should be roughly when the majority of the excitation has transitioned out of $EW(r)$.

Using purity of the global state, we can write this difference of relative entropies as the full modular energy
\begin{align}
S(\rho^{\psi_{\lambda}}_r|\sigma_r)- S(\rho^{\psi_{\lambda}}_{LI}|\sigma_{LI})
 = 2\pi\braket{H_r}_{\psi_{\lambda}} -2\pi \braket{H_{LI}}_{\psi_{\lambda}} = -\braket{\log \Delta_{\Omega;r}}_{\psi_{\lambda}}
\end{align}
where we have used the definition of the ``full'' modular operator,
\begin{align}
\Delta_{\Omega;r} = \sigma_r \otimes \sigma_{LI}^{-1} \equiv e^{-2\pi H_r} \otimes e^{2\pi H_{LI}},
\end{align}
in terms of the ``half'' modular Hamiltonians $H_r,\ H_{LI}$. As we will see, the full modular energy is much simpler to compute than the fidelity, so we will start there. Computing the full modular energy will require understanding the modular Hamiltonian for $LI$, $H_{LI}$. In general, finding explicit formulae for $H_{LI}$ is prohibitively difficult. We will find that progress can be made in the large distance limit, $\dl/r_I\ll 1$. 

Before turning to computing $H_{LI}$ in the large distance limit, we describe a toy model in two dimensions which we use to more explicitly illustrate our formulae and compute exact results.

\subsection{Solvable toy model in $\text{AdS}_2$}
As we will show, the information measures introduced in the previous subsection can be directly related to (integrals of) out-of-time-ordered correlators, similar to those considered in \cite{MSY-2}. We will be interested in studying the non-zero string length corrections to these correlators.  In higher dimensions, computing such OTOCs can be complicated by the presence of a direction transverse to the axis of scattering. In order to ignore this subtlety and make some of our computations more tractable, we consider a model first described in \cite{AMMZ}. 

We consider a theory of JT gravity coupled to a generalized free theory in $\text{AdS}_2$. We look at the theory at finite temperature and place it in equilibrium with a bath. The metric in the bath will be that of flat space. There will be transparent boundary conditions between the $\text{AdS}_2$ region and the flat space bath region. As discussed in \cite{AMM,AMMZ}, we should think of this description as a coarse grained description. The true microscopic description is given by two 0+1 dimensional quantum systems coupled to a 1+1 dimensional flat space theory. We imagine that the total system is in the thermofield double. Following the convention of \cite{AMM}, we take the action of the gravitational system to be
\begin{align}
    I = \frac{1}{4\pi} \int d^2 x \sqrt{-g} \left( \phi R + 2(\phi - \phi_0)\right) + I_{CFT}
\end{align}
where $\phi$ is the dilaton and $\phi_0$ is the extremal entropy of the black hole. Here $I_{CFT}$ will be the action of $K$ decoupled CFTs. We take the limit that $1/G_N\gg K \gg 1$ in order to enhance quantum effects due to bulk entanglement. 

\begin{figure}
 \centering
 \includegraphics[width = .8\textwidth]{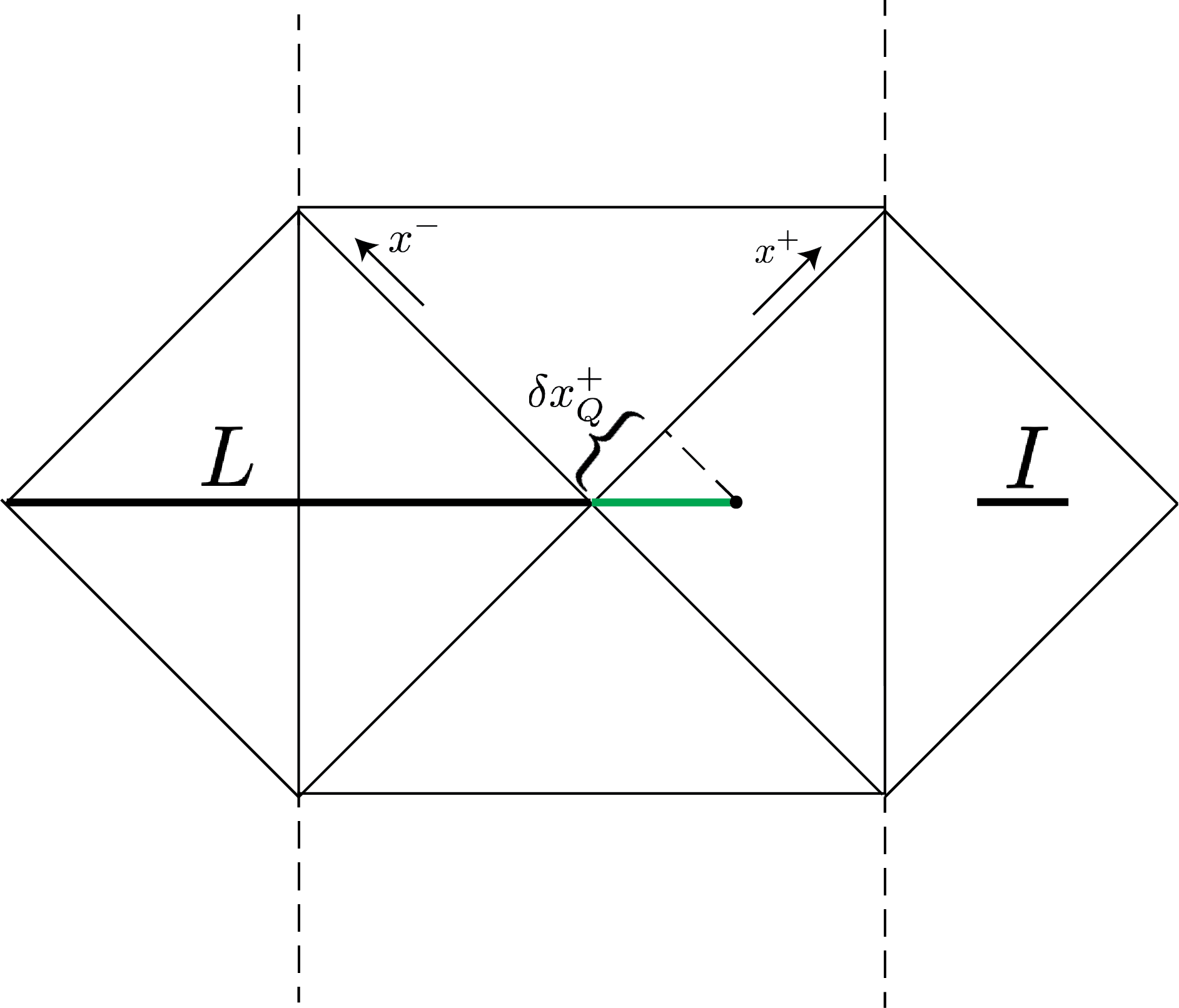}
 \caption{Our setup in a model with JT gravity coupled to a 1+1 dimensional CFT, which is also coupled to an external bath.}
 \label{fig:ads2regions}
\end{figure}

In this case, the region $LI$ will be given by the union of the whole left bath plus quantum system  $-~L~-$  and a small interval $I$ in the right, flat space bath, as in Figure \ref{fig:ads2regions}. As discussed in \cite{AMMZ}, there is a corresponding quantum extremal surface in the coarse grained description (the ``bulk'' dual to the microscopic description). This surface then lies in the exterior of the $\text{AdS}_2$ black hole closest to the interval $I$. We consider the metric in the $\text{AdS}_2$ region to be
\begin{align}
ds^2 = - \frac{4dx^+ dx^-}{(x^+x^-+1)^2},
\end{align}
where we have set $\ell_{\text{AdS}} = 1$.

The dilaton profile takes the form\footnote{Note that there are two ``$G_N$'''s in JT gravity. One $G_N$, inversely proportional to the extremal entropy, controls the suppression of higher topologies. The other controls the coupling to the Schwarzian mode of bulk fields. The $G_N$ we use here is the latter. In the notation of \cite{AMM}, $G_N = \frac{\beta}{2\pi \phi_r}$.}
\begin{align}
\phi(x^+,x^-) = \phi_0 + \frac{4\pi}{G_N} \frac{1-x^+ x^-}{1+x^+x^-}.
\end{align}
The position of the quantum extremal surface for $LI$ can be found by extremizing the generalized entropy functional \begin{align}\label{eqn:Sgen}
S_{gen}(e(LI)) = \phi(x^+_Q,x^-_Q) + S_{bulk}(x_Q^+, x_Q^-)
\end{align}
where $e(LI)$ is a region in the bulk which includes the whole left black hole exterior (plus bath) and a bit in the right exterior, see Figure \ref{fig:ads2regions}. The endpoints of $e(LI)$ in right exterior lie at $x^{\pm} = x_Q^{\pm}$. The quantum extremal surface can be found by extremizing over all the endpoints of $e(LI)$ which do not lie in the bath region.

Because the generalized entropy involves a bulk entropy term, it is quite hard to compute in general, especially when the region is disconnected. Thankfully we can make progress in the limit where $I$ is small and the bulk matter is given by a CFT, with conformal dimension $\delta$ for the lightest operator in its spectrum.  In that case, the answer was computed in \cite{Agon:2015ftl} and is given by 
\begin{align}
S_{bulk}(e(LI)) = - \frac{\sqrt{\pi} \Gamma(2\Delta_O +1)}{2^{4\Delta_O +2}\Gamma(2\Delta_O + 3/2)} \left( \frac{\dl}{r_I}\right)^{2\Delta_O} 
\end{align}
where $\dl$ is the length of the interval, $I$, $r_I$ is the null-coordinate distance from the quantum extremal surface to $I$, and $\Delta_O$ is the conformal dimension of the field. Note that we will be working here with purely right-moving fields in the bulk. Our results can be easily extended to non-chiral fields.  

In order to well separate the quantum extremal surface from the horizon, we imagine that there are $K$ flavors of this field which all contribute to support the quantum extremal surface. For simplicity, we assume that $1\ll K \ll 1/G_N$. In that case, the quantum extremal surface for $LI$ lies close to the bifurcation surface at $x^+ = x^-=0$. Extremizing \eqref{eqn:Sgen} with respect to $x^-$ gives us an equation for $x^+$. We find that the surface lies at 
\begin{align}\label{eqn:QESposition}
\delta x^+_Q = \frac{1}{4\pi} G_N K \frac{\sqrt{\pi}\Delta_O \Gamma(2\Delta_O +1)}{2^{4\Delta_O+2}\Gamma(2\Delta_O + 3/2)} \frac{\dl^{2\Delta_O} }{r_I^{2\Delta_O +1}}.
\end{align}

To probe the quantum extremal surface position, we need to scatter strings in the bulk. In the eikonal limit, the gravitational scattering matrix is given just by a phase $e^{i\delta_2(s)}$ with
\begin{align}
e^{i\delta_2(s)}= e^{iG_N p_+ q_-},
\end{align}
where $p_+q_-$ is the center of mass energy of the collision. In string theory, this scattering phase gets modified \cite{Shenker:2014tu,Brower:2007uc}. We model a stringy bulk theory by using the scattering ``phase''
\begin{align}\label{eqn:scatteringphase}
e^{i\delta_J(s)} = e^{-G_N(-ip_+q_-)^{J-1}}.
\end{align}
Note that we put ``phase'' in quotes because this amplitude has magnitude less than one. As discussed in \cite{Shenker:2014tu,Brower:2007uc}, this has to do with inelastic effects in string scattering. We will see that the imaginary component of $\delta_J$ is key to preserving the necessary causality conditions in the correlation functions we consider.

With this model in mind, we now turn to understanding the corrections to the modular Hamiltonian for $H_{LI}$ in the small $I$ limit.


\section{Modular Hamiltonian in the Large Distance Limit}\label{sec:3}

In this section, we use the replica trick to compute the form of the vacuum modular Hamiltonian for the region $LI$. We will summarize the main results, leaving the details to Appendix \ref{app:modham}. To make the problem tractable in general, we work to leading order in the ``large-distance limit,'' where the region $I$ is a sphere far from $L$, relative to $I$'s size. The basic idea is to extend the work of \cite{Agon:2015ftl}. In a 1+1 dimensional CFT, this large distance limit is quantified by taking the cross ratio associated to the four endpoints of the intervals $L$ and $I$ to be small. See figure \ref{fig:regions3dparticle} for a reminder of the setup.

Our strategy will be to find a formula for the expectation value of $H_{LI}$ in a dense set of states and extract the form of the modular Hamiltonian from this expectation value. We can compute the expectation value of the modular Hamiltonian via a replica trick, using the formula 
\begin{align}\label{eqn:reptrick}
\braket{H_{LI}}_{\psi} = -\frac{1}{2\pi} \partial_n \vert_{n=1} \text{Tr}[\psi_L \rho_{LI} \psi_L^{\dagger} \rho_{LI}^{n-1}],
\end{align}
where the $\psi_L$ are state creation operators inserted some amount into Euclidean time. We will attempt to compute $ \text{Tr}[\psi_L \rho_{LI} \psi_L^{\dagger} \rho_{LI}^{n-1}]$ at integer $n$ and then continue in $n$.

This calculation can be done by computing correlation functions involving twist operators, which are operators in the $CFT^{\otimes n}$ theory that can be thought of as living at the boundaries of $L$ and $I$ and which implement the twisted boundary conditions. As before, we take the small $I$ limit so that we can do the OPE of the twist operator(s) at $I$'s boundary. Denoting the $I$ twist operators by $\Sigma_n^I$ and the diameter of $I$ by $\dl$, we have the formula
\begin{align}
\Sigma_n^I \sim \braket{\Sigma_n^I}_{\Omega^{\otimes n}} \left( 1+ (\dl)^{2\Delta_O} \sum_{j\neq k}^n c_{j-k} O^{(j)} O^{(k)} + o(\dl^{2\Delta_O}) \right)
\end{align}
where $O^{(j)} = 1 \otimes ... O \otimes...1$, with the $O$ inserted in the $j$'th copy of the CFT. The analytic continuation in $n$ of the coefficients $c_{j-k}$ were computed by Agon \& Faulkner around $n=1$ in any dimension and are fixed by conformal symmetry. Note that there are no contributions from single copy operators because $\braket{\Sigma_n^I O^{(j)}}$ can be computed by conformally transforming to hyperbolic space, where the one-point function vanishes by symmetry. Furthermore, the OPE coefficients $c_{j-k}$ depend only on the difference $j-k$ because of replica symmetry.

Plugging this ansatz into equation \eqref{eqn:reptrick}, we find that we need to compute
\begin{align}\label{eqn:reptricksum}
&\frac{\text{Tr}[\rho_{LI}^{n-1} \psi \rho_{LI} \psi^{\dagger}]}{\text{Tr}[\rho_{LI}^n]} \approx \frac{\text{Tr}[\rho_L^{n-1} \psi\rho_L \psi^{\dagger}]}{\text{Tr}[\rho_L^n]} +(\dl)^{2\Delta_O} \sum_{k=0}^{n-2} c_k \text{Tr}[\rho_L^{n-1} O^{(k)} \psi_L \rho_L^{1/2} O \rho_L^{1/2} \psi^{\dagger}]\nonumber \\
&+(\dl)^{2\Delta_O} \sum_{j=0}^{n-3} \sum_{k=1}^{n-2-j} c_k \text{Tr}[ \rho_L^{n-1}O^{(j+k)} O^{(j)}\psi_L \rho_L\psi_L^{\dagger}]
\end{align}
where $O^{(j)} \equiv \rho_L^{-j} O_I \rho_L^{j} = \rho_L^{-j-1/2}O \rho_L^{j+1/2}$. Now we just need to analytically continue the latter two terms in $n$. We follow the method in \cite{Agon:2015ftl,Faulkner:2014aa} wherein we rewrite the sums as contour integrals. We focus on the first of the two sums.

\subsection*{Analytically continuing the first sum}
Following Agon \& Faulkner, we write
\begin{align}\label{eqn:firstsum}
&\sum_{k=0}^{n-2} c_k \text{Tr}[\rho_L^{n-1} O^{(k)} \psi_L \rho_L^{1/2} O \rho_L^{1/2} \psi^{\dagger}] \nonumber \\
& = \oint_{\mathcal{C}} ds \,k_n(s)\,c_n(-is) \text{Tr}[\rho_{L}^{n-1}O( -is+\pi) \psi_L \rho_L^{1/2} O \rho_L^{1/2} \psi^{\dagger}]
\end{align}
where
\begin{align}
O(-is) \equiv \rho_L^{is/2\pi} O \rho_L^{-is/2\pi}
\end{align}
and 
\begin{align}
k_n(s+2\pi i) = \frac{1}{2\pi i} \left(\frac{1}{e^s-1} -\frac{1}{e^{s/n+2\pi i/n}-1}\right)
\end{align}
and where the contour $\mathcal{C}$ circles the poles at $\text{Im} s = 2\pi i k$ for $k=0,..., n-2$. Note that the second term in $k_n(s)$ does not have poles at any of these values of $k$. We have added it in for later convenience, following the strategy in \cite{Agon:2015ftl}.

The OPE coefficients $c_n(-is)$ were computed in \cite{Agon:2015ftl} and are equal to a hyperbolic space thermal two-point function. Other than the thermal periodicity condition, $c_n(-is+2\pi n-\epsilon) = c_n(-is+\epsilon)$, we will not need the explicit form of $c_n$ away from $n=1$. At $n=1$, they take the form
\begin{align}\label{eqn:c1}
c_1(-is+\pi) = \frac{1}{(2\cosh(s/2))^{2\Delta_O}}.
\end{align}
Note that this formula holds for any spacetime dimension $d$ of the CFT. Below, we specialize to $d=2$ and resum a particular set of descendant contributions to $\delta H_{LI}$. For this we will need more information about $c_n$.

To analytically continue the sum in $\eqref{eqn:firstsum}$, we unwrap the contour $\mathcal{C}$ and we will be left with two contributions, one along the $\text{Im}s = 2\pi (n-2) + \epsilon$ line and the other along the $\text{Im}s = -\epsilon$ line. We can drop any contributions at infinity \cite{Faulkner:2014aa}. After deforming the contour, we choose to set any phases of the form $e^{2\pi i n}$ in the denominators of $k_n$ to $1$. This is a particular choice in analytic continuation away from integer $n$, but it is the choice that has worked consistently in several other papers \cite{Agon:2015ftl,Faulkner:2014aa}. After making this choice, everything else is now writable in terms of thermal correlators, for which the analytic continuation in $n$ is just a continuation in (inverse) temperature. For this, as always, we pick the obvious continuation provided by continuing in temperature.

Having done this, we can shift the lower contour at $\text{Im}s = -\epsilon$ down by $\pi$ without hitting a branch cut since $ \psi_L$ and $O$ are separated by an angle of $\pi$ ($O$ is inserted in the $I$ region). We can also shift the upper contour by $+i\pi$. Taking the $n \to 1$ limit, we find that the two contours cancel off each other to give
\begin{align}\label{eqn:firstsumfinal}
&\sum_{k=0}^{n-2} c_k \text{Tr}[\rho_L^{n-1} O^{(k)} \psi_L \rho_L^{1/2} O \rho_L^{1/2} \psi^{\dagger}] \nonumber \\
& \sim (n-1) \int_{-\infty}^{\infty} ds \frac{c_1(-is+\pi)}{4\cosh^2(s/2)}  \text{Tr}[\rho_L \rho_L^{-1/2} O \rho_L^{1/2}\psi^{\dagger}O (-is)\psi_L] + \mO((n-1)^2) \nonumber \\
&  = (n-1) \int_{-\infty}^{\infty} ds \frac{\braket{\psi_L^{\dagger} O_I O_L(-is) \psi_L}}{(2\cosh(s/2))^{2\Delta_O +2}} + \mO((n-1)^2).
\end{align}
where we used \eqref{eqn:c1} to plug in for $c_1$.

\subsection*{Second Sum}

For the second sum, we follow the same procedure. The details can be found in Appendix \ref{app:modham}. The answer is 
\begin{align}\label{eqn:secsumfinal}
&\sum_{j=0}^{n-3} \sum_{k=1}^{n-2-j} c_n(2\pi k) \text{Tr}[ \rho_L^{n-1}O^{(k+j)} O^{(j)}\psi_L \rho_L\psi_L^{\dagger}]\nonumber \\
&= \frac{(n-1)}{2\pi i} \int \frac{ds_j ds_k}{4\cosh^2(s_j/2)} \left(\frac{1}{e^{s_k+i\epsilon}-1} + \frac{1}{e^{s_k+s_j}+1}\right) \times \nonumber \\
&  c_1(-is_k+\epsilon)\times \Braket{\psi^{\dagger}O_L(-is_k -is_j) O_L(-is_j) \psi}+ \mO((n-1)^2).
\end{align}

Plugging \eqref{eqn:secsumfinal} and \eqref{eqn:firstsumfinal} into \eqref{eqn:reptricksum}, we find the operator equation
\begin{align}
\delta H = &\frac{-(\dl)^{2\Delta_O}}{2\pi}\int_{-\infty}^{\infty} \frac{1}{(2\cosh(s/2))^{2\Delta_O +2}} \rho_L^{is} O_L(-r_I)\rho_L^{-is} O_I(r_I) \\
& \hspace{-.6cm} +\frac{i (\dl)^{2\Delta_O}}{4\pi^2 } \int \frac{ds_j ds_k}{4\cosh^2(s_j/2)} \left(\frac{1}{e^{s_k+i\epsilon}-1} + \frac{1}{e^{s_k+s_j}+1}\right)c_1(-is_k + \epsilon) O_L(-is_k -is_j) O_L(-is_j) \nonumber \\ \nonumber & \hspace{-.6cm} +\delta H_{II}
\end{align}
where $-r_I$ denotes the Rindler reflected position (i.e. $O(-r_I) = J_L O(r_I) J_L$. The final term, $\delta H_{II}$, is just the second term mapped under the conformal transformation which exchanges the two intervals, $L$ and $I$. We will discuss this transformation further below. The $\delta H_{II}$ term needs to be there since the full answer $H_{LI}$ needs to be invariant under this transformation. We do not see the contribution from $\delta H_{II}$ expanding about small $\dl$. The $\delta H_{LL}$ and $\delta H_{II}$ terms will not affect most of our analysis below and so we do not focus on them.

\subsection{Contributions from Descendants}
When we compute the fidelity in Section \ref{sec:6}, we need to also include contributions to the modular Hamiltonian from a certain class of descendants. In particular, we can write the twist operator OPE as
\begin{align}
\sigma_{-n} \sigma_n \sim \braket{\sigma_{-n} \sigma_n} \left( 1+ (\dl)^{2\Delta_O} \sum_{j\neq k}^n \sum_{n,m=0}^{\infty} (\dl)^{n+m} c^{n,m}_{j,k} \partial^n O^{(j)} \partial^m O^{(k)}+... \right)
\end{align}
where the operators $\partial^n O$ are located at the center of the interval $I$.

As we did for the non-descendant primaries $O$, we can extract these coefficients $c_{j,k}^{n,m}$ by computing the overlap
\begin{align}\label{eqn:expansion}
&\braket{\sigma_{-n} \sigma_n O^{(j)}(z) O^{(k)}(z')}  \sim \braket{\sigma_{-n} \sigma_n} \times \nonumber \\
&\left( 1+ (\dl)^{2\Delta_O} \sum_{a,b=0}^{\infty} (\dl)^{a+b} c^{a,b}_{j,k} \braket{\partial^aO^{(j)}(0) O^{(j)}(z)} \braket{\partial^b O^{(k)}(0) O^{(k)}(z')} +... \right) \nonumber \\
& =  \braket{\sigma_{-n} \sigma_n} \times  (\dl)^{2\Delta_O} \sum_{a,b=0}^{\infty} (\dl)^{a+b} c^{a,b}_{j,k} \frac{(-1)^{a+b} \Gamma(2\Delta_O+1)^2}{\Gamma(2\Delta_O +1-n)\Gamma(2\Delta_O + 1-m)} \frac{1}{z^{2\Delta_O+a} (z')^{2\Delta_O+b}} + ... \nonumber \\
\end{align}
where for simplicity we have moved the center of the interval $I$ to $z= \bar{z} = 0$. We see that we can extract the coefficients $c_{j,k}^{a,b}$ by expanding $\braket{\sigma_n \sigma_n O^{(j)}(z) O^{(k)}(z)}$ as a double power series in small $\dl/z, \,\dl/z'$. In Appendix \ref{app:modham}, we compute the coefficients $c_{j,k}^{a=0,b}$ for general $b$. We find
\begin{align}\label{eqn:descopecoeff}
c_{j,0}^{a=0,b} = \frac{1}{\left(e^{\pi i j/n} - e^{-i\pi j /n}\right)^{2\Delta_O}}\left( \frac{e^{\pi i j/n} + e^{-i\pi j/n}}{e^{\pi i j/n}- e^{-i\pi j/n}}\right)^b \frac{1}{b! \,2^b \,n^{b+2\Delta_O}} + \text{(terms proportional to $n-1$)}.
\end{align}

We can account for this infinite sum of descendant contributions to $\delta H_{LI}$ by analytically continuing 
\begin{align}
\sum_{k=0}^{n-2} \sum_b c_k^{a=0,b} \text{Tr}[\rho_L^{n-1} O^{(k)} \psi_L \rho_L^{1/2} \partial^b O \rho_L^{1/2} \psi^{\dagger}].
\end{align}
Running through the same steps as detailed after equation \eqref{eqn:firstsum} above, we find that the modular Hamiltonian takes the form
\begin{align}\label{eqn:nonresummed}
\delta H_{LI} = \frac{-1}{8\pi}(\dl/2)^{2\Delta_O}\int ds \frac{1}{\cosh^{2\Delta_O+2}(s/2)} \times \left( \sum_{b=0}^{\infty} \frac{1}{b!} (\dl/2)^m \tanh^b(s/2) \times O_L(re^s) \partial^b O_I(-r) \right).
\end{align}
One can neatly resum these contributions by writing 
\begin{align}\label{eqn:resummed}
\delta H_{LI} = -\frac{1}{2\pi} (\dl)^{2\Delta_O}\int ds \frac{1}{(2\cosh(s/2))^{2\Delta_O+2}} \times O_L(re^s)  O_I(\overline{re^s})
\end{align}
where
\begin{align}
\overline{x^-} = -r_- \frac{x^- +r_+}{x^- + r_-},\ \ r_{\pm} = r \pm \dl/2.
\end{align}
This map is a Mobius transformation that exchanges the interval $I$ with the Rindler wedge $L$. To see that \eqref{eqn:resummed} gives \eqref{eqn:nonresummed}, we have expanded $O_I(\overline{re^s})$ in $\dl/r_I$ and dropped terms that are higher order in $\dl$ (i.e. terms of the form $\dl^{a} \partial^b O_I$ where $a>b$). This form of $\delta H_{LI}$ will be used below in Section \ref{sec:5}.


\section{Computing the Full Modular Energy}\label{sec:4}
We turn now to computing the difference in relative entropies
\begin{align}
S(\psi_{\lambda} | \Omega;r) - S(\psi_{\lambda} | \Omega;LI) = 2\pi\braket{H_r}_{\psi_{\lambda}} - 2\pi \braket{H_{LI}}_{\psi_{\lambda}} = -\braket{\log \Delta_r}_{\psi_{\lambda}}
\end{align} 
where $H_r,\,H_{LI}$ are the vacuum Hamiltonians for the regions $r$ and $LI$ respectively. We aim to prove our main result \eqref{eqn:stringysdiff} by utilizing our knowledge of $H_{LI}$ in the small interval limit. 

Since $\ket{\psi_{\lambda}}$ is just unitary evolution of the vacuum by the Pauli's defined in \eqref{eqn:pauli}, we can write
\begin{align}
-\braket{\log \Delta_r}_{\psi_{\lambda}} = - 2\pi Z_{\delta}^2 \sin^2(\lambda) \Braket{\phi_R(T_R-i\delta) \left( H_r - H_{LI}\right)\phi_R(T_R+i\delta)}_{\Omega},
\end{align}
where $Z_{\delta}$ is the overall normalization for the state $\phi_R(T_R+i\delta)\ket{\Omega}$ which is $Z_{\delta} = (2\sin(\delta))^{\Delta_{\phi}}$.
We can compute this expectation value by analytically continuing the correlator
\begin{align}\label{eqn:Hcomm}
\mathcal{M} (t_L, t_R)= - 2\pi \,Z_{\delta}^2 \sin^2(\lambda) \Braket{\phi_L(t_L) \left( H_r - H_{LI}\right)\phi_R(t_R)}_{\Omega} = -2\pi Z_{\delta}^2 \sin^2(\lambda) \braket{[H_{LI},\phi_L]\phi_R}_{\Omega}
\end{align}
to $t_L \to -T_R -i\pi+i\delta$ from real $t_L$ and $t_R \to T_R+i\delta$. In the second equality, we have used that $[H_r,\phi_L]=0$ since $r$ is contained in $R = \overline{L}$ together with the equality $H_r \ket{\Omega} = H_{LI}\ket{\Omega}$.

In the previous section, we found that\footnote{Note that when we plug in for $H_{LI}$ from the previous section, strictly speaking we are using the equivalence between bulk and boundary modular Hamiltonians in the 2-d setup in \ref{sec:2}. The ``boundary'' region $L$ is the whole left bath together with the left boundary quantum mechanical system. The ``bulk'' region $L$, to the order we work, can effectively be taken to just be the whole left black hole exterior together with the bath. We use the same symbol for both since this distinction will be unimportant for us.}
\begin{align}
H_{LI} = H_L + H_I + \delta H_{LI} + \delta H_{LL} + \delta H_{II},
\end{align}
where the $\delta H$ operators are bilocal in the lightest operator $O$ in the CFT. Plugging this into \eqref{eqn:Hcomm}, we see that 
\begin{align}\label{eqn:Hcommnew}
\mathcal{M} (t_L, t_R)/(Z_{\delta}^2 \sin^2(\lambda))= -2\pi \braket{[H_L + \delta H_{LI},\phi_L]\phi_R}_{\Omega} -2\pi\braket{[ \delta H_{LL},\phi_L]\phi_R}_{\Omega} .
\end{align}
Assuming that we do not pick $\phi$ to be the same operator as $O$, this final term is suppressed by $G_N$. Furthermore, unlike the first two terms, this final term does not grow like $e^{-T_R}$ as $T_R$ gets large and negative and so will just result in a $G_N$-suppressed shift in the time $T_R^*$ at which $-\braket{\log \Delta_r}$ transitions from positive to negative. We thus ignore this term.

To compute the remaining terms, we plug in for $\delta H_{LI}$ using \eqref{eqn:resummed} and find that $\mathcal{M}$ is given by the sum of a two-point function and an out-of-time-order commutator
\begin{align}\label{eqn:timederiv}
\mathcal{M} (t_L, t_R)/(Z_{\delta}^2 \sin^2(\lambda))  = 2\pi i \braket{\left(\partial_{t_L} \phi_L\right) \phi_R} - 2\pi \int_L dx f(x) \braket{[O^i_L(x) O^i_I(\bar{x}),\phi_L]\phi_R}.
\end{align}

Four-point functions of just this form have been computed in \cite{MSY-2}. As just discussed, we are looking for a term which is growing exponentially with $T_R$. The only ordering that produces this growing piece is
\begin{align}
-2\pi \int_L dx f(x) \braket{\phi_LO_L^i O_I^i\phi_R}.
\end{align}
As discussed in \cite{MSY-2,Shenker:2014tu}, this can be computed via a bulk scattering amplitude. Working in momentum-space, we can expand the boundary operators in the momentum basis and integrate the wavefunctions against the scattering phase in \eqref{eqn:scatteringphase}. One then gets
\begin{align}
&-2\pi  \int_L dx f(x) \braket{\phi_LO_L^i O_I^i\phi_R} \nonumber \\
&= -2\pi K\int dp_+ \braket{\phi_L(e^{t_L})|p_+}\braket{p_+|\phi_R(e^{t_R})} \int dx f(x) \int dq_- \braket{O^i_R(\overline{x})|q_-}\braket{q_-|O^i_L(x)}e^{i \delta_J (p_+ q_-)}
\end{align}
where the scattering phase $i \delta_J(p_+ q_-) = -G_N(-ip_+q)^{J-1}$ and the momentum wavefunctions are given by 
\begin{align}
&\braket{\phi_L(e^{t_L})|p_+}\braket{p_+|\phi_R(e^{t_R})} =  \frac{e^{\Delta_{\phi}(-t_L+t_R)}}{\Gamma(2\Delta_{\phi})} \frac{(2ip_+)^{2\Delta_{\phi}}}{(-p_+)}e^{-i2(e^{-t_L}+e^{t_R})p_+}\theta(-p_+) ,\nonumber \\
&\braket{O^i_R(x^- =\overline{re^s})|q_-}\braket{q_-|O^i_L(x^-=re^{s})}  =  \frac{e^{\Delta_O s}}{\Gamma(2\Delta_O)} \frac{(2iq_-)^{2\Delta_O}}{(-q_-)}e^{-i2r(e^{s}+1)q_-}\theta(-q_-).
\end{align}

We will be interested in the probe limit, where $\phi$ does not backreact significantly on the geometry. This limit can be implemented by expanding the exponential scattering phase in small $\delta _J$. At leading order in this expansion, we find a term which factorizes. This is just the Wick contraction of the $\phi$'s and $O$'s amongst themselves. This term will be canceled by a corresponding term in the other ordering of the four-point function in \eqref{eqn:timederiv}. At next order, we find a term which grows exponentially with $t_L - t_R$:
\begin{align}\label{eqn:growing}
&-2\pi  \int_L dx f(x) \braket{\phi_LO_L^i O_I^i\phi_R} \nonumber \\ 
& = - \int dp_+ \braket{\phi_L(e^{t_L})|p_+}\braket{p_+|\phi_R(e^{t_R})} (-p_+)^{J-1} \delta \tilde{x}_Q^{(J)} + (\text{term which cancels out})
\end{align}
with
\begin{align}\label{eqn:dxq}
&\delta \tilde{x}_Q^{(J)} = (\dl)^{2\Delta_O}G_N K\int_{-\infty}^{\infty} ds \frac{1}{(2\cosh(s/2))^{2\Delta_O+2}} \int dq_- \braket{O_R(\overline{re^s})|q_-}\braket{q_-|O_L(re^{s})}(i q_-)^{J-1} \nonumber \\
&  = (\dl)^{2\Delta_O}G_N K\frac{C(\Delta_O,J)}{r_I^{2\Delta_O + J-1}}\frac{\sqrt{\pi}\Gamma (2 \Delta_O +1) \Gamma (J+2 \Delta_O )}{2^{2\Delta_O+1}\Gamma(2\Delta_O + \frac{J}{2} + \frac{1}{2}) \Gamma(2\Delta_O + \frac{J}{2} + 1)}
\end{align}
where
\begin{align}
C(\Delta_O, J) = \frac{\Gamma(2\Delta_O + J-1)}{\Gamma(2\Delta_O)4^{\Delta_O + (J-1)}}.
\end{align}

The $p_+$ integral in \eqref{eqn:growing} can be done explicitly and we find
\begin{align}\label{eqn:fullmodmainresult}
-\braket{\log \Delta_r}_{\psi_{\lambda}} = \frac{2\pi \sin^2(\lambda) \Delta_{\phi}}{\sin(\delta)}\left( \cos(\delta) - \delta x_Q^{(J)} e^{-(J-1)T_R} \sin^{2-J}(\delta)+ \mathcal{O}((G_N)^2)\right)
\end{align}
where
\begin{align}\label{eqn:xshift}
\delta x_Q^{(J)} \equiv \frac{4^{\Delta_{\phi}}C(\Delta_{\phi},J)}{2\pi \Delta_{\phi}} \delta \tilde{x}_Q^{(J)}.
\end{align}

For small $\delta$, this precisely matches the form in \eqref{eqn:stringysdiff}. Equation \eqref{eqn:fullmodmainresult} is the main result of this section and one of the main results of this paper. We now reiterate a few comments made in the introduction:
\begin{itemize}
\item As can be checked, for $J = 2$, the time at which this quantity switches from positive to negative, $T_R^*(J=2)$, is independent of $\Delta_{\phi}$ and only weakly dependent on the smearing scale $\delta$.
\item For $1< J<2$, $T_R^*(J)$ is larger in magnitude than in the case of gravity, $|T_R^*(J)| > |T_R^*(2)|$.
\item For $1< J <2$, the time $T_R^*(J)$ depends on the smearing scale as $T_R^*(J) \supset (2-J) \log (\sin(\delta))$ which diverges with a smaller smearing scale. Of course, we should keep in mind that for too small a smearing scale backreaction effects of the excitation will become important. 
\end{itemize}

The full modular Hamiltonian is equal to the difference in relative entropies between the region $r$ and its complement, $LI = \overline{r}$. Intuitively, when the relative entropy between the excited state and the vacuum is large in $r$, it should be small in $LI$ leading to a positive difference in relative entropies. This is because when the relative entropy in $r$ is large, this means the density matrices are more easily distinguishable in $r$. This means the excitation in the state $\psi$ must be more well-localized in $r$'s entanglement wedge than in $LI$'s.

Thus, one might want to interpret $T_R^*(J)$ as roughly the time at which the excitation in $\ket{\psi}$ crosses from being mostly in $r$'s entanglement wedge to being mostly in $LI$'s.  Indeed, for $J=2$, the turnover time is at 
\begin{align}
e^{T^*_R(J=2)} = \delta x_Q^{(2)}.
\end{align} 
Using equation \eqref{eqn:dxq}, comparing with equation \eqref{eqn:QESposition} and remembering that for near right-boundary operators $x^+ = e^{T_R}$, we see that the full modular energy transitions from positive to negative precisely when the operator is null-separated from the quantum extremal surface. Thus, this interpretation appears to check out in the case with bulk gravity, $J=2$. Unfortunately, we can be less sure for $J<2$, since there is no rigorous reason why the difference in the relative entropies we have computed needs to encode whether the string is reconstructable from $LI$ or $r$.

The quantity that \emph{does} rigorously encode which region the string's state is reconstructable from is the quantum fidelity susceptibility, which we turn to estimating in the next two sections.


\section{Computing modular flowed correlators in the Regge limit}\label{sec:5}
In the previous section we computed the difference in relative entropies, which served as an intuitive measure of when the perturbation transitions from the entanglement wedge of $LI$ to that of $r$. However, as discussed in the introduction, it is the fidelity \eqref{eqn:fidelity} that rigorously encodes which entanglement wedge the perturbation resides in. Therefore, our ultimate goal is to compute it using perturbation theory. As will become clear in the next section, at leading order in $\lambda$ the fidelity can be obtained from the correlator 
\begin{align}\label{mainquant}
\mathcal{F}(t_L, t_R; s) \equiv \langle \phi_L(t_L) \Delta^{is/2\pi}_L \Delta^{-is/2\pi}_{LI}\phi_R(t_R)\rangle. 
\end{align} 

In this section we compute $\mathcal{F}(t_L, t_R;s)$ in the Regge limit, which we take to mean the limit of large $t_L-t_R$, before turning to the fidelity in the next section. As in previous sections, we work to leading order in the size of the interval, $\dl$. To proceed, note that we can write \eqref{mainquant} in terms of density matrices as 
\begin{align}\label{eqn:Fexpansion}
\mathcal{F}(s)  = \langle \rho^{is/2\pi}_{LI}\rho^{-is/2\pi}_L \phi_L\rho^{is/2\pi}_L \rho^{-is/2\pi}_{LI}\phi_R\rangle 
\end{align}
where we have temporarily suppressed the kinematics of the state insertions to avoid clutter. In this form we can readily expand $\mathcal{F}(s)$ perturbatively in $\delta H_{LI}$ using Baker-Campbell-Hausdorff (BCH). At leading order in $\dl$, the expression is 
\begin{align}\label{BCH}
\mathcal{F}(s) = & \braket{\phi_L \phi_R} + i\int_{0}^s dt' \langle [\phi_L, \delta H_{LI}(t')]\phi_R\rangle + \int_0^s dt dt' \langle \delta H_{LI}(t)\phi_L \delta H_{LI}(t')\phi_R\rangle \nonumber \\
& - \frac{1}{2}\int_0^s dt dt' \langle \overline{\mathcal{T}} \left( \delta H_{LI}(t) \delta H_{LI}(t')\right) \phi_L\phi_R\rangle - \frac{1}{2}\int_0^s dt dt' \langle\phi_L \mathcal{T} \left( \delta H_{LI}(t) \delta H_{LI}(t')\right) \phi_R\rangle 
\end{align}
where $\delta H_{LI}(t) = (\rho_L \otimes \rho_I)^{it/2\pi}\delta H_{LI}(\rho_L \otimes \rho_I)^{-it/2\pi}$ and $\mathcal{T}$ and $\overline{\mathcal{T}}$ denote time-ordering and anti time-ordering with respect to evolution by $ (\rho_L \otimes \rho_I)^{it/2\pi}$, respectively. In the end, the terms of the form $\braket{\delta H \delta H \phi \phi}$ and $\braket{\phi \delta H \delta H \phi}$ will not grow in the Regge limit, so we can ignore them for our purposes. 

The form of the modular Hamiltonian that we found in Section \ref{sec:3} is\footnote{We will drop the indices on the $O$'s and implicitly sum over all $K$ flavors.} 
\begin{align}\label{modham5}
\delta H_{LI} = -\frac{1}{2\pi}(\dl)^{2\Delta_O}\int ds \frac{1}{(2\cosh (s/2))^{2\Delta_O +2}}O_L(r_I e^s)O_I( \overline{r_I e^{s}}).
\end{align}
At face value it would appear as if equation \eqref{BCH} includes terms subleading in $\dl$ by going to second order in the BCH formula \eqref{BCH}. However, from \eqref{modham5}, note that the second order term in BCH will contain, at leading order in $1/N$, a Wick contraction between the two $O_I$ insertions. Since we are working perturbatively in $\dl$, this contraction generates an enhancement $1/(\dl)^{2\Delta_O}$, essentially due to both operators being inserted within the same small interval. Thus the second term in \eqref{BCH} actually contributes at the same order as the naive leading term.\footnote{One might worry that for this reason we need to include an infinite number of terms from BCH. One can check that the third term in \eqref{BCH} is the only one which receives a large enough $1/\dl$ enhancement to compete with the second term in \eqref{BCH}.} 

In order to proceed, we break up the computation of \eqref{BCH} into three parts. At leading order in $1/N$, both terms in \eqref{BCH} involve four-point functions analogous to those in Section \ref{sec:4}. Therefore, we compute these in the Regge limit in the same way by mapping them onto a bulk scattering process. We allow for the exchange of operators of spin $1 < J \leq 2$ between the scattered particles. In Section \ref{linord} we first compute \eqref{BCH} to linear order in $G_N$. In Section \ref{realitycond} we show that the result satisfies an important consistency condition that is interesting in its own right. Finally in Section \ref{exponenGN} we show that the result exponentiates in terms of $G_N$, when working at large $K$. 

\subsection{Computing $\mathcal{F}(s)$ at linear order in $G_N$}\label{linord}
We start by working out $\delta H_{LI}(t)$. The $\rho_L$ factor simply corresponds to a Rindler time translation of $O_L$. The $\rho_I$ factor does the same to $O_I$, up to an overall conformal factor:
\begin{align}\label{eqn:conformalfactor}
\rho_I^{i t}O_I(\overline{r_I e^{s}})\rho_I^{-i t} = \Omega^{\Delta_O}(s,t)O_I(\overline{r_I e^{s-t}}),
\end{align}
where
\begin{align}
\Omega(s,t) = \frac{\cosh^2(s/2)}{\cosh^2((s+t)/2)}.
\end{align}
See Appendix \ref{app:conformalfactor} for more details.

Let us now denote the first and second terms in \eqref{BCH} by $\mathcal{F}_1(s)$ and $\mathcal{F}_2(s)$, respectively. To begin with, note that both $\mathcal{F}_1$ and $\mathcal{F}_2$ contain multiple operator orderings. However, since we are working in the Regge limit we only need to keep the growing terms, which corresponds to the ordering $\langle \phi_L O_L O_R \phi_R \rangle$. 

In the Regge limit, the first term is easily evaluated: 
\begin{align}
\mathcal{F}_1(s) &\sim i\int_0^s dt \langle \phi_L \delta H(t) \phi_R \rangle \nonumber 
\\&= -\frac{i}{2\pi 2^{2\Delta_O +2}}(\dl)^{2\Delta_O}\int_0^s dt \int \frac{ds'}{\cosh^{2\Delta_O +2}(s'/2)}\frac{\cosh^{2\Delta_O}(s'/2)}{\cosh^{2\Delta_O}((s'-t)/2)}\langle \phi_L O_L(r_I e^{s'-t})O_I(\overline{r_I e^{s'-t}})\phi_R \rangle \nonumber 
\\&\approx \mathcal{O}(G_N^0) + \frac{i}{2^{2\Delta_O+2}2\pi}KG_N e^{-(J-1)T_R} \frac{(\dl)^{2\Delta_O}}{r_I^{2\Delta_O + J-1}} C(\Delta_O,J)C(\Delta_{\phi},J)e^{i\pi(J-1)/2} \nonumber 
\\&\times\int_0^s dt \int \frac{ds'}{\cosh^2(s'/2)}\frac{e^{(J-1)(-s' + t)/2}}{\cosh^{4\Delta_O+J-1}((s'-t)/2)}
\end{align}  
where 
\begin{align}
C(\Delta_O, J) = \frac{\Gamma(2\Delta_O + J-1)}{\Gamma(2\Delta_O)4^{\Delta_O + (J-1)}}.
\end{align}
By shifting $s' \rightarrow s' + t$ and doing the $t$ integral we are left with 
\begin{align}
\mathcal{F}_1(s) \sim -\frac{1}{2^{2\Delta_O+2}2\pi }KG_N e^{-(J-1)T_R}\frac{(\dl)^{2\Delta_O}}{r_I^{2\Delta_O + J-1}} C(\Delta_O,J)C(\Delta_{\phi},J)\int ds' \frac{g_J(s',s)}{\cosh^{4\Delta_O+J-1}(s'/2)},
\end{align}
where we define 
\begin{align}\label{eqn:gJ}
g_J(s',s) = 2i e^{i\pi (J-1)/2}e^{-(J-1)s'/2}\left(\tanh(s'/2)-\tanh\left(\frac{s'+s}{2}\right) \right).
\end{align}

Now we evaluate the second term in the Regge limit. To start with we have 
\begin{align}
\mathcal{F}_2(s) &= \frac{(\dl)^{4\Delta_O}}{(4\pi)^2}\int_0^s dt dt' \int \frac{ds ds'}{(2\cosh(s/2))^{2\Delta_O +2}(2\cosh(s'/2))^{2\Delta_O +2}} \nonumber 
\\ &\times \left(\Omega(s,t)\Omega(s',t') \right)^{\Delta_O}\langle O_L(r_I e^{s-t})O_I(\overline{r_I e^{s-t}})\phi_L O_L(r_I e^{s'-t})O_I(\overline{r_I e^{s'-t}})\phi_R\rangle 
\end{align}
Naively this looks higher order in $\dl$ than the first term. However, note that at leading order in $1/N$ one of the possible Wick contractions is between the two $O_I$ insertions:  
\begin{align}\label{intermedcalc}
\mathcal{F}_2(s) &\sim \frac{(\dl)^{4\Delta_O}}{(4\pi)^2}\int_0^s dt dt' \int \frac{ds ds'}{(2\cosh(s/2))^{2\Delta_O +2}(2\cosh(s'/2))^{2\Delta_O +2}} \nonumber 
\\ &\times \left(\Omega(s,t)\Omega(s',t') \right)^{\Delta_O}\langle O_I(\overline{r_I e^{s-t}})O_I(\overline{r_I e^{s'-t}}) \rangle\langle O_L(r_I e^{s-t})\phi_L O_L(r_I e^{s'-t})\phi_R\rangle 
\end{align}
The two-point function can be computed from the results in Appendix \ref{app:conformalfactor}: 
\begin{align}
\langle O_I(\overline{r_I e^{s-t}})O_I(\overline{r_I e^{s'-t}}) \rangle = (\dl)^{-2\Delta_O}e^{-i\pi \Delta_O}2^{2\Delta_O}\frac{\cosh^{2\Delta_O}((s-t)/2)\cosh^{2\Delta_O}((s'-t')/2)}{\sinh^{2\Delta_O}((s-s'-t+t'-i\epsilon)/2)}. 
\end{align}
The important point is that this diverges like $(\dl)^{-2\Delta_O}$ which arises from the fact that both $O_I$ insertions are in the same small interval and hence contain a leading UV divergence as $\dl \rightarrow 0$. 

Now, the remaining four-point function in \eqref{intermedcalc} can be evaluated in the Regge limit by analytically continuing from the growing correlator $\braket{\phi_L O_L O_R\phi_R}$ without crossing any branch cuts, 
\begin{align}
\langle O_L(r_I e^{s-t})\phi_L O_L(r_I e^{s'-t})\phi_R\rangle  = \langle O_R(r_I e^{s-t-\tilde{t}})\phi_L O_L(r_I e^{s'-t})\phi_R\rangle \Big\lvert_{\tilde{t}\rightarrow -i\pi},
\end{align}
which can be evaluated as before in terms of bulk scattering. The result is 
\begin{align}
\mathcal{F}_2(s) &\approx \mathcal{O}(G_N^0) - \frac{1}{4\pi^2 2^{2\Delta_O +4}}KG_N e^{-(J-1)T_R}\frac{(\dl)^{2\Delta_O}}{r_I^{2\Delta_O + J-1}} C(\Delta_O,J)C(\Delta_{\phi},J) e^{2i\pi \Delta_O + i\pi (J-1)/2}\nonumber 
\\ &\times \int_0^s dt dt' \int \frac{ds ds'}{\cosh^2(s/2)\cosh^2(s'/2)}\frac{e^{(J-1)(-s-s'+t+t')/2}}{\sinh^{4\Delta_O + J-1}((s'+t-s-t'+i\epsilon)/2)} 
\end{align}
We now perform a series of contour manipulations. First we shift $s \rightarrow s + t, ~ s' \rightarrow s' + t'$, and evaluate the $t$ integrals. Then we shift $s' \rightarrow s' + s$, which allows us to evaluate the $s$ integral. We are left with the following expression: 
\begin{align}
\mathcal{F}_2(s) \approx -\frac{1}{\pi 2^{2\Delta_O+4}}KG_N e^{-(J-1)T_R}\frac{(\dl)^{2\Delta_O}}{r_I^{2\Delta_O + J-1}} C(\Delta_O,J)C(\Delta_{\phi},J)\int ds' \frac{f_J(s',s)}{\cosh^{4\Delta_O + J-1}(s'/2)}
\end{align}
where
\begin{align}\label{eqn:fJ}
f_J(s',s) &= e^{(J-1)s/2}\sinh(s/2) \nonumber 
\\ &\times \frac{\left( -2\cosh(Js'/2)\sinh((J-2)s/2) - \sinh(J(s-s')/2 + s') + \sinh(s' - J(s+s')/2) \right)}{\cos(\pi J/2)\cosh(s'/2)\cosh((s'+s)/2)\cosh((s-s')/2)}
\end{align}

Summing the two terms in \eqref{BCH} we find
\begin{align}\label{eqn:finalleadingorder}
\mathcal{F}(s) = -\frac{1}{2^{2\Delta_O +2}2\pi}KG_N e^{-(J-1)T_R}\frac{(\dl)^{2\Delta_O}}{r_I^{2\Delta_O + J-1}} C(\Delta_O,J)C(\Delta_{\phi},J)\int ds'\frac{ \left(g_J(s',s) + \frac{1}{2}f_J(s',s) \right)}{\cosh^{4\Delta_O + J-1}(s'/2)},
\end{align}
where we remind the reader that we have dropped the terms that order $G_N^0$. We will add them back in the next subsections.

This answer has a few interesting features. First, one can check that $g_J + \frac{1}{2} f_J$ goes to zero at $s=0$. At large $s$, this quantity takes the form
\begin{align}
g_J(s',s) + \frac{1}{2}f_J(s',s) \sim e^{(J-1)s} C(s').
\end{align}
which grows exponentially but at a slower rate given by the different Lyapunov exponent. It was shown in \cite{Boer:2019td, Faulkner:2018vl} that correlation functions of the form \eqref{eqn:Fexpansion} obey a ``modular-chaos'' bound, which states that the growth of the correlator in $s$ cannot exceed $e^s$. Here we find an example of a correlation function which does not saturate the bound. In this case, the modular chaos Lyapunov exponent just inherits the value of the ``regular'' Lyapunov exponent.

\subsection{Reality condition}\label{realitycond}
Equation \eqref{eqn:finalleadingorder} passes multiple checks, one of which we now discuss. Consider the correlator 
\begin{align}
\mathcal{G}(t_L, t_R;s) = \langle \phi_L(t_L)\Delta_{LI}^{-is/2\pi}\Delta_L^{is/2\pi}\phi_R(t_R)\rangle 
\end{align}
This is related to the correlator in \eqref{mainquant} by 
\begin{align}
\mathcal{G}(t_L, t_R; s) = \mathcal{F}(t_L-s, t_R+s;s). 
\end{align}
There is a non-trivial consistency condition which $\mathcal{G}(t_L, t_R;s)$ must satisfy. Namley,  $\mathcal{G}(t_L, t_R; s+i\pi)$ must be real. This property follows from the fact that this correlator can be written in terms of a map which acts as an inclusion, which we now explain. Consider the map\footnote{This map coincides with the twirled Petz map under an inclusion \cite{OhyaPetz}. This map is defined respect to a vector (the vacuum in our setup) that is jointly cyclic and separating for the two regions.} 
\begin{align}
    \chi_s(\phi_R) = \Delta_r^{is/2\pi}J_r J_R \Delta_r^{-is/2\pi}\phi_R \Delta_r^{is/2\pi}J_R J_r \Delta_r^{-is/2\pi}.
\end{align}
Since the modular conjugation operator $J$ maps observables in one algebra to its commutant, one can check that $\chi_s(\phi_R) \in \mA_r$ if $\phi_R \in \mA_R$. A simple computation shows that  
\begin{align}
    \langle \phi_L(t_L) \chi_s(\phi_R(t_R))\rangle = \langle \phi_L(t_L) \Delta_{LI}^{-i(s + i\pi)/2\pi}\Delta_L^{i(s +\pi)/2\pi}\phi_R(t_R)\rangle = \mathcal{G}(t_L, t_R; s+i\pi)
\end{align}
where we have used both \eqref{tomitaidentity} and \eqref{commutantidentity}. Now, 
\begin{align}
    2i \text{Im}\langle \phi_L(t_L) \chi_s(\phi_R(t_R))\rangle = \langle [\phi_L(t_L),\chi_s(\phi_R(t_R))]\rangle 
\end{align}
but note that $\chi_s(\phi_R) \in \mathcal{A}_r$ so this commutator vanishes, hence the reality condition follows.  

We now show that the reality condition is indeed satisfied using the results from the previous section. We start by noting that 
\begin{align}
\mathcal{G}(t_L, t_R;s+i\pi) = \mathcal{O}(G^0_N) - \frac{KG_N e^{-(J-1)(t_R-t_L)} (\dl/r_I)^{2\Delta_O}C(\Delta_O)C(\Delta_{\phi})}{2^{2\Delta_O+2}\cosh^{2\Delta_{\phi}+J-1}(\frac{t_L + t_R}{2})} \int ds' \frac{G_J(s',s)}{\cosh^{4\Delta_O + J-1}(s'/2)} 
\end{align} 
where 
\begin{align}
G_J(s',s) &= -\frac{e^{-(J-1)s/2}\cosh(s/2)}{2} \nonumber 
\\ &\times \frac{\left( -2\cosh(Js'/2)\cosh((J-2)s/2) + \cosh(J(s-s')/2 + s') + \cosh(s' - J(s+s')/2) \right)}{\cos(\pi J/2)\sinh((s-s'-i\epsilon)/2)\sinh((s+s'-i\epsilon)/2)\cosh(s/2)}.
\end{align}
Now, in the $s'$ plane this expression has poles at $s' = -s + i\epsilon$ and $s' = s - i\epsilon$. Evaluating the residue at this pole yields a vanishing result as $\epsilon \rightarrow 0$. Therefore, we can separately do the following: push the $s'$ contour up by $i\pi - i\epsilon$ or push it down by $-i\pi + i\epsilon$. Since these two expressions are just contour deformations of the original integral which don't cross any poles with non-zero residue, we can just add the two expressions and divide by two to get back $\mathcal{G}(t_L, t_R;s+i\pi)$. Doing so, we find that the lower contour simply yields the complex conjugate of the upper contour, hence $\mathcal{G}(t_L, t_R; s+i\pi)$ is indeed a real function. 

\subsection{Exponentiation in $G_N$}\label{exponenGN}
    Thus far we have been working to linear order in $G_N$, expanding the scattering phase $e^{i \delta_J(p_+q_-)} \approx 1+i\delta_J(p_+q_-)$. As explained in Section \ref{sec:2}, we ultimately want to work in the large $K$ limit. We now show that in this limit, the linear order result we derived in Section \ref{linord} actually exponentiates. To this aim, we return back to the correlator we were computing in Section \ref{linord}, i.e. 
\begin{align}
\mF(s) = \braket{\phi_L \Delta_L^{is} \Delta_{LI}^{-is} \phi_R} = \braket{\rho_{LI}^{is} \rho_L^{-is} \phi_L \rho_L^{is} \rho_{LI}^{-is} \phi_R}.
\end{align}
Now, if we expand $H_{LI} = H_L + H_I + \delta H_{LI}$, then using BCH, we have that 
\begin{align}
\rho_{LI}^{-is} = \rho_L^{-is} \otimes \rho_I^{-is} \mathcal{T} \exp \left( i \int_0^s dt' \delta H(t') \right)
\end{align}
where
\begin{align}
\delta H(t') = (\rho_L \otimes \rho_I)^{it'} \delta H (\rho_L \otimes \rho_I)^{-it'},
\end{align}
and the time-ordering symbol is with respect to time generated by $\rho_L \otimes \rho_I$.
Thus, we just get the correlator
\begin{align}
\mF(s) = \Braket{\overline{\mathcal{T}} \left(e^{-i \int_0^s dt' \delta H_{LI}(t')}\right) \phi_L \mathcal{T}\left( e^{i\int_0^s dt' \delta H_{LI}(t')}\right) \phi_R},
\end{align}
where $\overline{\mathcal{T}}$ denotes anti time-ordering.

Now, we can expand these exponentials and then begin computing the correlators at each order using Wick contractions. The only contractions which will actually notice the time-ordering symbols are contractions involving two $\delta H_{LI}$ operators. Accounting for this, we get
\begin{align}
\mF(s) =&\int dp_+ \braket{\phi_L|p_+}\braket{p_+|\phi_R} \times \nonumber \\
&\sum_{m,n=0}^{\infty} \frac{(-i)^m i^n}{m! n!} \sum_{\substack{k_{\alpha} \\ k_1 + k_2 + 2k_4 = n \\k_1+k_3 + 2k_5 = m}} \left(\wick{\braket{ \delta \c1 H \phi_L(p_+)  \delta \c1 H \phi_R}}\right)^{k_1}\left(\wick{\braket{\phi_L \c1 \delta  \c1 H \phi_R(p_+)}}\right)^{k_2} \times \nonumber \\
&  \braket{\delta H}^{k_3} \mathcal{T} \left(\braket{\delta H \delta H}^{k_4}\right) \overline{\mathcal{T}} \left(\braket{ \delta H \delta H}^{k_5}\right) \times C_{k_{\alpha}}(n,m) \nonumber \\
& + (\text{terms which do not grow with $e^{-T_R}$ or are down in $K$ counting})
\end{align}
where we have abbreviated notation so that $\delta H \equiv \int_0^s dt \delta H_{LI}(t)$. Furthermore, we have defined the various factors
\begin{align}
\left(\wick{\braket{ \delta \c1 H \phi_L(p_+)  \delta \c1 H \phi_R}}\right)^{k_1} &\equiv \prod_{i=1}^{k_1} \int dq_-^i e^{-G_N(-i p_+ q_-^i)^{J-1}} \nonumber 
\\ &\times \int dx_1^i dx_2^i f_{LI}(x_1^i)f_{LI}(x_2^i) \braket{O_L(x_1^i)|q^i_-}\braket{q_-^i| O_L(x_2^i)} \braket{O_I(\overline{x_1^i}) O_I(\overline{x_2^i})} \nonumber \\
\left(\wick{\braket{\phi_L \c1 \delta  \c1 H \phi_R(p_+)}}\right)^{k_2} &\equiv \prod_{i=1}^{k_2} \int dq_-^i e^{-G_N(-i p_+ q_-^i)^{J-1}} \int dx^i f_{LI}(x^i) \braket{O_L(x^i)|q_-^i}\braket{q_-^i| O_I(\overline{x^i})} 
\end{align}
where the $f_{LI}$ are placeholders for the smearing function in \eqref{modham5} together with the integral from $0$ to $s$ in the defintion of $\delta H$. The coefficients $C_{k_{\alpha}}(n,m)$ are combinatorial factors, counting all the ways of contracting various operators. We find
\begin{align}
C_{k_{\alpha}}(n,m)= {n \choose k_1, k_2} {m \choose k_1, k_3} k_1 ! (2k_4 -1)!! (2k_5-1)!!
\end{align}
where the multinomial factors are for choosing the $k_1$, $k_2$ and $k_3$ factors from $n$ and $m$. The $k_1!$ counts the number of ways to contract the $\delta H$'s. The double factorials are for all the ways of contracting the remaining $2k_{4,5}$ factors of $\delta H$ amongst themselves. Note that we have also dropped terms which have fewer $K$-index loops, since these will suppresed in the $K \to \infty$ limit. Using that $(2k-1)!! =(2k)!/(k! 2^k)$, we get 
\begin{align}\label{eqn:BCHexpansion}
\mF(s) =\int dp_+  \braket{\phi_L|p_+}\braket{p_+|\phi_R} &\sum_{m,n=0}^{\infty} \sum_{\substack{k_{\alpha} \\ k_1 + k_2 + 2k_4 = n \\k_1+k_3 + 2k_5 = m}} \frac{\left(\wick{\braket{ \delta \c1 H \phi_L(p_+)  \delta \c1 H \phi_R}}\right)^{k_1}}{(k_1)!}\frac{\left(\wick{\braket{\phi_L \c1 \delta  \c1 H \phi_R(p_+)}}\right)^{k_2}}{(k_2)!}\times \nonumber \\
& \frac{(-i)^{k_3}\braket{\delta H}^{k_3}}{(k_3)!}\mathcal{T} \left(\frac{(-1)^{k_4} \braket{\delta H \delta H}^{k_4}}{2^{k_4}(k_4)!}\right) \overline{\mathcal{T}} \left(\frac{(-1)^{k_5}\braket{ \delta H \delta H}^{k_5}}{2^{k_5} (k_5)!}\right),
\end{align}
where we have moved the time-ordering symbol into the sum to act on the only factors which are actually affected by the time-ordering symbol. One can show that that 
\begin{align}
    \sum_{k=0}^{\infty}\mathcal{T} \left(\frac{(-1)^{k_4}\braket{ \delta H \delta H}^{k_4}}{2^{k_4} (k_4)!}\right) = e^{-\frac{1}{2}\mathcal{T}\braket{\delta H \delta H}},
\end{align}
along with the analogous formula for the anti time-ordered factors in \eqref{eqn:BCHexpansion}. We see that all the factors in \eqref{eqn:BCHexpansion} exponentiate and we just get
\begin{align}
\mF(s) = \int dp_+\braket{\phi_L|p_+}\braket{p_+|\phi_R} \exp[-i \braket{\delta H} - \braket{\delta H\delta H}] \times \exp \left[\wick{\braket{ \delta \c1 H \phi_L(p_+)  \delta \c1 H \phi_R}} + i\wick{\braket{\phi_L \c1 \delta  \c1 H \phi_R(p_+)}} \right],
\end{align}
where we have also used the fact that $\frac{1}{2} \left(\mathcal{T}+ \overline{\mathcal{T}}\right) \braket{\delta H \delta H} = \braket{\delta H \delta H}$.

In the end, we can write the final expression as 
\begin{align}\label{mainresult}
\mathcal{F}(s) = \int dp_+\braket{\phi_L|p_+}\braket{p_+|\phi_R} \exp\left[F_J(s)(-p)^{J-1} \right] 
\end{align} 
where $F_J(s)$ is essentially what we obtained in Section \ref{linord}, modulo the $p_+$ integral: 
\begin{align}\label{eqn:FullFJ}
F_J(s) = -\frac{e^{-i\pi (J-1)/2}}{2\pi 2^{2\Delta_O +2}}KG_N \frac{(\dl)^{2\Delta_O}}{r_I^{2\Delta_O + J-1}}C(\Delta_O,J)\int ds' \frac{g_J(s',s) + \frac{1}{2}f_J(s',s)}{\cosh^{4\Delta_O + J-1}(s'/2)},
\end{align}
with $g_J,\,f_J$ given in \eqref{eqn:gJ}, \eqref{eqn:fJ} respectively. This is the main result of this section. Armed with this expression, in the next section we describe how to use \eqref{mainresult} to obtain the fidelity, and compute it for various operator dimensions and spins. 

\subsection{A Check for $J=2$}
The correlator in \eqref{mainquant} was recently computed in maximally chaotic ($J=2$) CFTs in \cite{Levine:2020wz}. The answer found there was that $F_J(s) \sim 2i (e^s-1) \delta x_Q^+$ where $\delta x_Q^+$ is the null shift of the quantum extremal surface for $LI$ from including $I$. For $J=2$, our function $F_J$ in \eqref{eqn:FullFJ} takes the form
\begin{align}\label{eqn:F2}
F_2(s) = \frac{i}{2\pi 2^{4\Delta_O +3}}G_N K \frac{\Delta_O(\dl)^{2\Delta_O}}{r_I^{2\Delta_O + 1}}\int ds' \frac{\left(g_2(s',s) + \frac{1}{2}f_2(s',s)\right)^+}{\cosh^{4\Delta_O + 1}(s'/2)}
\end{align}
where $\left(g_2(s',s) + \frac{1}{2}f_2(s',s)\right)^+$ is $g_2(s',s) + \frac{1}{2}f_2(s',s)$ but symmetrized under $s' \to -s'$ since it is integrated against an even function. Plugging in equations \eqref{eqn:gJ} and \eqref{eqn:fJ}, we find 
\begin{align}
\left(g_2(s',s) + \frac{1}{2}f_2(s',s)\right)^+= \frac{e^s-1}{\cosh(s'/2)}.
\end{align}
Thus, we get
\begin{align}\label{eqn:F2final}
F_2(s) = \frac{i}{2\pi 2^{4\Delta_O +2}}G_N K \frac{\Delta_O(\dl)^{2\Delta_O}}{r_I^{2\Delta_O + 1}} \frac{\sqrt{\pi} \Gamma(2\Delta_O+1)}{\Gamma(2\Delta_O + 3/2)} (e^s-1) =  2i\delta x_Q^+ (e^s-1)
\end{align}
where $\delta x_Q^+$ is the null position of the quantum extremal surface, given in equation \eqref{eqn:QESposition}. Thus, we find agreement with \cite{Levine:2020wz}.


\section{Computing the fidelity}\label{sec:6}
We now turn to computing the fidelity susceptibility. As discussed in the introduction, the fidelity between $\ket{\psi_{\lambda}^{X,Y}} = e^{i\lambda \sigma_{X,Y}}\ket{\Omega}$ and $\ket{\Omega}$ directly measures how well one can reconstruct the excitation from $LI$. The fidelity only depends on the density matrices of $\psi$ and $\Omega$ on $r$. One can derive an explicit formula
\begin{align}
-2 \log F(\psi,\Omega; r) = \text{Tr} \left[\sqrt{\sqrt{\sigma_r^{\Omega}} \rho_r^{\psi} \sqrt{\sigma_r^{\Omega}}}\right] \equiv S_{1/2, 1/2} (\rho^{\psi}_r | \sigma^{\Omega}_r).
\end{align}
where $S_{\alpha, z}(\rho | \sigma) = $ is the \emph{$\alpha -z$ relative entropy}. 
For small $\lambda$, our states $\ket{\psi_{\lambda}^{X,Y}}$ are perturbatively close to the vacuum, so we know that 
\begin{align}
\rho_r^{\psi_{\lambda}} = \sigma_r^{\Omega} + \lambda \delta \rho.
\end{align}
We can then expand the above expression for the fidelity to leading order in $\lambda$. As discussed in the introduction, the fidelity takes the form
\begin{align}
F(\psi_{\lambda}, \Omega; r) = 1- \frac{\lambda^2}{2} \chi(\psi,\Omega;r) + \mathcal{O}(\lambda^3)
\end{align}
with $\chi$ the fidelity susceptibility.

The fidelity susceptibility was computed in \cite{May:2018ti} where the authors found
\begin{align}
&\chi(\psi,\Omega;r) = \frac{1}{2} \frac{d^2}{d\lambda^2} \left(S_{1/2, 1/2}(\rho^{\psi_{\lambda}}_r| \sigma_r) \right) \nonumber \\
& = \int ds \tilde{P}_{1/2,1/2}(s) \text{Tr}[\sigma^{-1} \delta \rho \sigma^{-is/2\pi} \delta \rho \sigma^{is/2\pi}]
\end{align}
where $\tilde{P}_{1/2,1/2}(s)$ is the Fourier transform of the function
\begin{align}
P_{1/2,1/2}(\omega) = \frac{1}{1+ e^{2\pi \omega}}.
\end{align}

For states of the form
\begin{align}\label{eqn:statesevolvedbyr}
\ket{\psi} = (1+i\lambda \Delta_r^{\delta/2\pi}O_r) \ket{\Omega},
\end{align}
we show in Appendix \ref{app:susceptibility} that the susceptibility takes the form
\begin{align}\label{eqn:fidelityoperatorinsertion}
\chi(\psi, \Omega;r) =  \frac{-1}{2\pi}\text{Im} \int ds \left( \frac{\braket{\delta \psi| \Delta_r^{is/2\pi}|\delta \psi}}{\sinh((s+i\epsilon)/2)}- i \frac{\braket{\delta \psi| \Delta_r^{is/2\pi}J_r|\delta \psi}}{\cosh(s/2)}\right)
\end{align}
with 
\begin{align}
\ket{\delta \psi} \equiv i \Delta_r^{\delta/2\pi} O_r \ket{\Omega}.
\end{align}

Note that our states at small $\lambda$ take the form
\begin{align}\label{eqn:statesreal}
\ket{\psi^{X,Y}_{\lambda}} = \left (1+ i \lambda \, Z_{\delta} \Delta_R^{\delta/2\pi} \phi_R(T_R)\right)\ket{\Omega}
\end{align}
where $\lambda$ is pure real for $X$ and pure imaginary for $Y$. These states differ from the states in \eqref{eqn:statesevolvedbyr} since the Euclidean and Lorentzian time evolutions are with respect to the region $R$ and \emph{not} the region $r$. However, we want to compute the fidelity for $r$ and not $R$. To get such a formula, we can imagine starting our operators $\phi_R$ in $r$ (since $r \subset R$). We can smear the operators a bit in Lorentzian time so as to make the states normalizable, but only over times such that the smeared $\phi_R$ still lies in $r$. Then we can justifiably use \eqref{eqn:fidelityoperatorinsertion}. To compute the fidelity for the true states of interest, \eqref{eqn:statesreal}, we can imagine analytically continuing the fidelity in the position of the operator $\phi_R(T_R)$ out of the region $r$ while remaining in $R$. The question is then how to do this analytic continuation. We choose to analytically continue each term in \eqref{eqn:fidelityoperatorinsertion} in $T_R$ separately. We will see below that this gives us self consistent results (i.e. a real and positive fidelity susceptibility). Of course, this does not amount to a rigorous justification of the formulae we use. It would be interesting to prove equation \eqref{eqn:fidelityoperatorinsertion} for a more general class of $\ket{\delta \psi}$ that includes the states of interest for us.

Nevertheless, plugging our states in \eqref{eqn:statesreal} into equation \eqref{eqn:fidelityoperatorinsertion} and doing the aforementioned analytic continuation, we get the equation
\begin{align}\label{eqn:fidelitymain}
\chi(\psi^X, \Omega;r) =  -\frac{Z_{\delta}^2}{2\pi}\text{Im} \int ds \left( \frac{\braket{\phi_R(T_R-i\delta) \Delta_r^{is/2\pi} \phi_R(T_R+i\delta)}}{\sinh((s+i\epsilon)/2)}+i \frac{\braket{\phi_R(T_R-i\delta) \Delta_r^{is/2\pi}J_r \phi_R(T_R+i\delta)}}{\cosh(s/2)}\right)
\end{align}
where we have been careful to remember that $J_r i = -i J_r$ since $J_r$ is anti-unitary. The formula for $\chi(\psi^Y,\Omega;r)$ is the same as \eqref{eqn:fidelitymain} but with the sign on the second term reversed.

To proceed, we need to compute each term in \eqref{eqn:fidelitymain} separately. To do this, we utilize the results in the previous section. We notice that each term in \eqref{eqn:fidelitymain} can be recast in terms of the correlation function $\mF(t_L, t_R;s)$ defined in the previous section
\begin{align}
\mF(t_L, t_R;s) = \braket{\phi_L(t_L) \Delta_L^{is/2\pi} \Delta_{LI}^{-is/2\pi} \phi_R(t_R)}.
\end{align}
In particular, the first term takes the form
\begin{align}
&\braket{\phi_R(T_R-i\delta) \Delta_r^{is/2\pi} \phi_R(T_R+i\delta)} = \mF(t_L = -T_R-s-i\pi + i\delta, t_R = T_R+i\delta;s).
\end{align}
The second term can be computed by inserting $J_R^2 =1$ to the right of $\phi_R(T_R-i\delta)$. We can then act the $J_R$ on $\phi_R$ and use $\bra{\Omega}\phi_R(T_R-i\delta) J_R = \bra{\Omega}\phi_L(-T_R-i\delta)$. Finally, we will be left with the product $J_RJ_r = J_{L}J_{LI}$. This product acts on a dense set of states as $J_L J_{LI} = \Delta_L^{-1/2} \Delta_{LI}^{1/2}$. Using this, we see that the second term can be written as 
\begin{align}
&\braket{\phi_R(T_R-i\delta) \Delta_r^{is/2\pi} J_r\phi_R(T_R+i\delta)} = \braket{\phi_L(-T_R-i\delta) \Delta_L^{-1/2} \Delta_{LI}^{1/2-is/2\pi} \phi_R(T_R+i\delta)} \nonumber \\
& = \mF(t_L = -T_R-s-i\delta,t_R = T_R+i\delta; s+i\pi).
\end{align}

We can thus directly use the result \eqref{mainresult} of the previous section to compute the fidelity susceptibility in \eqref{eqn:fidelitymain}. To illustrate what we get, we begin with the case of gravity, $J=2$.

\subsection{The case of gravity, $J=2$} 
When $J=2$, we can plug equation \eqref{eqn:F2final} into \eqref{mainresult}. Using the form of the momentum wavefunction 
\begin{align}
\braket{\phi_L(t_L)|p_+}\braket{p_+|\phi_R(t_R)} =  \frac{e^{\Delta_{\phi}(-t_L+t_R)}}{\Gamma(2\Delta_{\phi})} \frac{(2ip_+)^{2\Delta_{\phi}}}{(-p_+)}e^{-i2(e^{-t_L}+e^{t_R})p_+}\theta(-p_+)
\end{align}
we have that the two terms in the fidelity take the form 
\begin{align}\label{eqn:twotermsJ2}
&\braket{\phi_R(T_R-i\delta) \Delta_r^{is/2\pi} \phi_R(T_R+i\delta)} \nonumber \\
&=\int_{-\infty}^0 dp_+ \frac{e^{i\pi \Delta_{\phi}} e^{\Delta_{\phi}(s+2T_R)}}{\Gamma(2\Delta_{\phi})} \frac{(2ip_+)^{2\Delta_{\phi}}}{(-p_+)}\exp\left(-i2e^{T_R}(e^{i\delta} - e^{s-i\delta})p_+ - 2i\delta x_Q^+ (e^s-1)p_+\right), \nonumber \\
& \braket{\phi_R(T_R-i\delta) \Delta_r^{is/2\pi}J_r \phi_R(T_R+i\delta)} \nonumber \\
& =\int_{-\infty}^0 dp_+ \frac{ e^{\Delta_{\phi}(s+2T_R+2i\delta)}}{\Gamma(2\Delta_{\phi})} \frac{(2ip_+)^{2\Delta_{\phi}}}{(-p_+)}\exp\left(-i2e^{T_R+i\delta}(e^s +1)p_+ + 2i\delta x_Q^+ (e^s+1)p_+\right)
\end{align}

Plugging these into \eqref{eqn:fidelitymain}, we can do a rough analysis of the behavior of the fidelity susceptibility as follows. First we know that in the strict limit $\delta \to 0$, the second term is perfectly well defined and finite whereas the first term diverges. To leading order in $\delta$ we can then ignore this term. We can now try to deform the $s$ contour on the first term in \eqref{eqn:fidelitymain} by $s \to s+i\pi$, avoiding the pole at $s = -i\epsilon$. If we can implement this contour deformation without crossing any non-analyticities in the complex s-plane, then we see from \eqref{eqn:twotermsJ2} that the first term just becomes order $\delta^0$ as well. A quick way to check whether this deformation is allowable is whether the real part of the exponential in the first line of \eqref{eqn:twotermsJ2} is negative for $s \to s+i\theta$ (with $\theta >0$). This ensures that the $p_+$ integral is convergent.

\begin{figure}
 \centering
 \includegraphics[width = \textwidth]{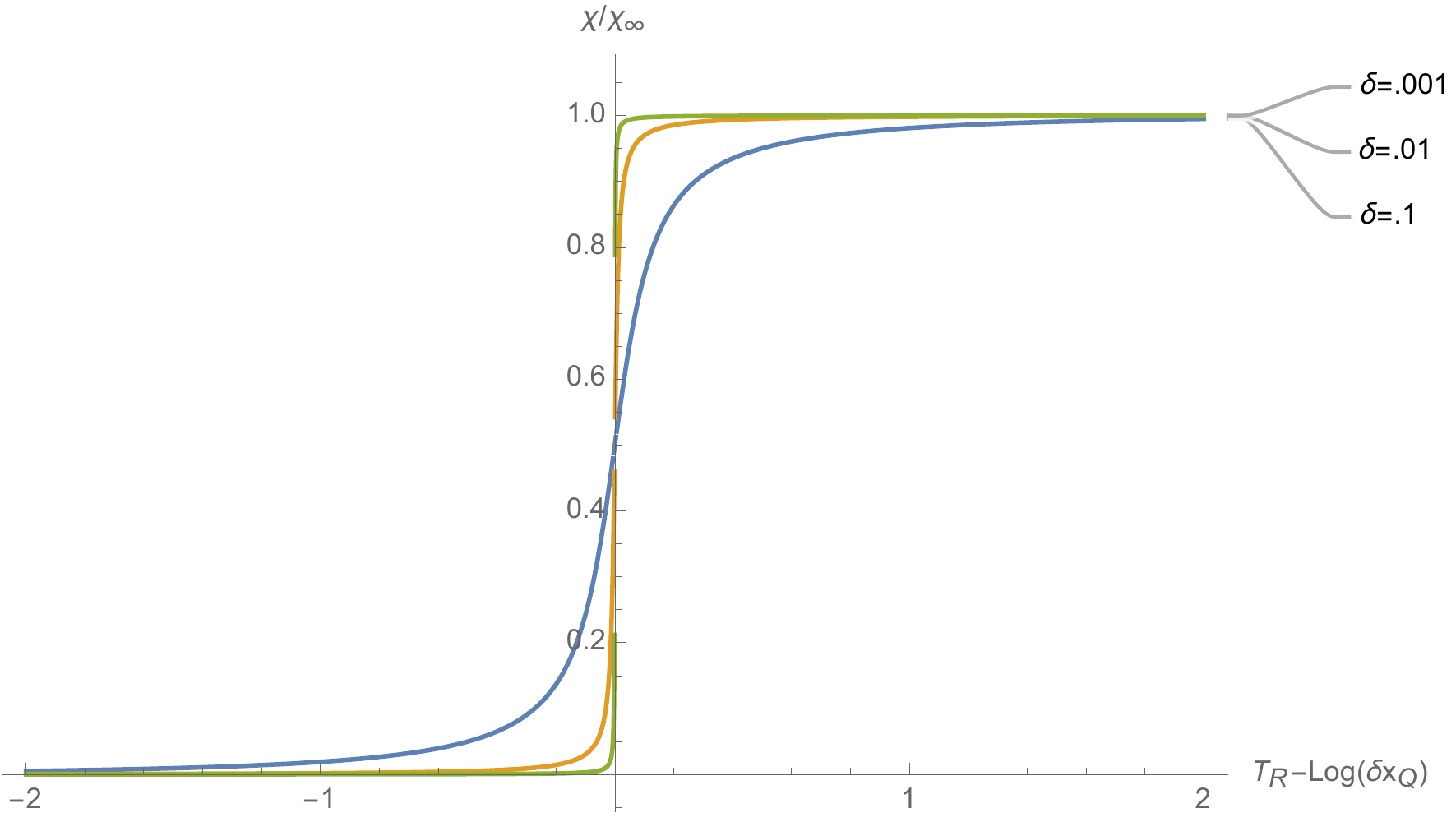}
  \caption{We plot the fidelity susceptibility normalized by its value at $T_R = \infty$ for $J=2$, $\Delta_{\phi}=1/2$ and $\Delta_O = 1/8$. Again, we have shifted the time axis so that the origin corresponds to $e^{T_R} = \delta x_Q$ where $\delta x_Q$ is the null position of the QES, computed in \eqref{eqn:QESposition}. The dependence of the turnover time on the smearing scale, $\delta$, has disappeared.}
  \label{fig:mainresultgrav}
\end{figure}

Thus, we need
\begin{align}
\text{Re}\left(-i2e^{T_R}(e^{i\delta} - e^{s-i\delta})p_+ - 2i\delta x_Q^+ (e^s-1)p_+\right) <0
\end{align}
which implies that 
\begin{align}
e^{T_R} (e^s \sin(\theta -\delta) - \sin(\delta)) < \sin(\theta)e^s \delta x_Q^+.
\end{align}
For small $\delta$, we see that this condition becomes just that $e^{T_R} < \delta x_Q^+$. If this condition is not obeyed, we can instead deform the $s$-contour for the first term in \eqref{eqn:fidelitymain} down in the complex $s$-plane so that the first term is order $\delta^0$ at small $\delta$. In other words, at small $\delta$, we have the formula
\begin{align}
\chi(\rho, \sigma) \approx  \Theta(e^{T_R} - \delta x_Q^+) + \mathcal{O}(\delta^{2\Delta_{\phi}})
\end{align}
where $\Theta(x) =1$ for $x>0$ and $\Theta(x)=0$ for $x<0$. The transition is sharp for small $\delta$. Futhermore, the transition time is independent of any features of the probe other than its coordinate position. This is what we expect for chiral operators probing a sharply defined quantum extremal surface. As a check of this result, we also have numerically computed the $s$-integrals in \eqref{eqn:fidelitymain} for $J=2$ and $\Delta =1/2$. We have ploted the results in Figure \ref{fig:mainresultgrav}.

\subsection{Saddle point analysis for $J<2$}
When the bulk string length is non-zero, we can analyze the integrals in \eqref{eqn:fidelitymain} via a saddle approximation. At small $\delta$, the dominant term is the first term of \eqref{eqn:fidelitymain} since it goes to infinity in the small $\delta$ limit. Indeed, 
\begin{align}
\chi(\psi^X, \Omega;r) \approx  -\frac{Z_{\delta}^2}{2\pi}\text{Im} \int ds \left( \frac{\braket{\phi_R(T_R-i\delta) \Delta_r^{is/2\pi} \phi_R(T_R+i\delta)}}{\sinh((s+i\epsilon)/2)}\right) + \mathcal{O}(\delta^0).
\end{align}

To simplify the computation and to compare with numerical results in the next subsection, we consider the case where the probe particle has $\Delta_{\phi} = 1/2$.
The integral we need to compute for this case is
\begin{align}
&\int_{-\infty}^{\infty} \frac{ds}{\sinh((s+i\epsilon)/2)}\braket{\phi_R(T_R-i\delta) \Delta_r^{is/2\pi} \phi_R(T_R+i\delta)} \nonumber \\
&=2\int_{-\infty}^{\infty} \frac{ds}{\sinh((s+i\epsilon)/2)}\int_{-\infty}^0 dp_+  \exp\left(4i \sinh(s/2 -i\delta)p_++ \tilde{F}_J(s) e^{-T_R(J-1)} (-p_+)^{J-1}\right)
\end{align}
with $\tilde{F}_J(s) = F_J(s) e^{-s(J-1)/2}$ and $F_J$ defined in \eqref{eqn:FullFJ}.

We can do the $p_+$ integral by saddle point. We find a solution at
\begin{align}
-p_+^*(s) = \left( \frac{(J-1)\tF_J(s)e^{-T_R(J-1)}}{4i\sinh(s/2-i\delta)}\right)^{\frac{1}{2-J}}.
\end{align}
Plugging this solution back in, the remaining $s$ integral becomes 
\begin{align}
& =2 \int_{-\infty}^{\infty} \frac{ds}{\sinh((s+i\epsilon)/2)} \frac{\exp\left(  \left( \frac{ (J-1)\tF_J(s)e^{-(J-1)T_R}}{(4i\sinh(s/2-i\delta))^{J-1}}\right)^{\frac{1}{2-J}} \left(\frac{2-J}{J-1} \right)\right)}{\sqrt{(J-2)(J-1)\tF_J(s)e^{-T_R(J-1)}(-p^*_+(s))^{J-3}}}.
\end{align}
Now, we would like to evaluate the $s$-integral via saddle point as well. Since we are working in the limit where $e^{-T_R}$ is large. We can do this by finding the saddle point solutions. One solution will be where the derivative of the coefficient of $e^{-T_R(J-1)/(2-J)}$ vanishes. This derivative is (including the terms in the denominator which we exponentiate)
\begin{align}
 &\left( \frac{ (J-1)\tF_J(s)e^{-T_R}}{(4i\sinh(s/2-i\delta))^{J-1}}\right)^{\frac{J-1}{2-J}} \left(\frac{\tF_J'(s)}{(4i\sinh(s/2-i\delta))^{J-1}} - (J-1)2i\cosh(s/2-i\delta) \frac{\tF_J(s)}{(4i\sinh(s/2-i\delta))^J}\right) \nonumber \\
 &-\frac{1}{2}\coth(s/2) + \frac{1}{2(2-J)} \frac{\tF'_J(s)}{\tF_J(s)} + \frac{J-3}{4(2-J)} \coth(s/2-i\delta) =0. 
\end{align}

Finding solutions to this equation is difficult in general since we do not have an analytic handle on $F_J$. We can try to find solutions at small $s$ as well as large positive and negative $s$. The latter two will be suppressed in the limit of small $\delta$ and so we can ignore them. We can look for solutions of the form $s \approx i\delta$. In this limit, the two terms in the parenthesis in the first line above need to vanish. The second line will just give corrections to the answer suppressed by $e^{T_R(J-1)/(2-J)}$. In this limit, we find a saddle at $s_* = \frac{2}{2-J} i \delta$.

Note, in order for our assumption that $s_*$ is small to hold, we need to always be taking the limit where $\delta \ll 2-J$. Furthermore, note that since $2-J <1$, then $s_*$ is above the singularity of the exponent in  at $s = 2i\delta$ in the complex plane. This requires us to deform the defining contour in \eqref{eqn:twotermsJ2} up and around the branch cut in the exponent. Presumably a more careful saddle point analysis would justify this point, but we just do a preliminary analysis here.

Plugging this saddle point back into \eqref{eqn:twotermsJ2} and accounting for the square root determinant, we get that the susceptibility is 
\begin{align}\label{eqn:saddlefinal}
\chi(\psi^{X,Y},\Omega;r) \approx 2\exp \left( -\left(\frac{\delta \tilde{x}_Q^{(J)} \delta^{2-J}e^{-(J-1)T_R}}{2\pi 2^{2J-3}}\right)^{\frac{1}{2-J}}\right) +\mathcal{O}(\delta),
\end{align}
where we have used \eqref{eqn:FullFJ} and \eqref{eqn:dxq} to write $|F'_J(0)|= \delta \tilde{x}_Q^{(J)}$. 

So far we have just computed the susceptibility for the region $r$. We can easily compute the fidelity for $LI$ as well just by sending $s \to -s$ in \eqref{eqn:fidelitymain}, where we are using that $\Delta_{LI}^{is} = \Delta_r^{-is}$. Since the second term in \eqref{eqn:fidelitymain} is order $\delta^0$, we see that the susceptibilities obey the equation
\begin{align}\label{eqn:sumrule}
    \chi_{LI} + \chi_r = 2 + \mathcal{O}(\delta),
\end{align}
where we have picked up the pole in the $s$ integral at $s= -i\epsilon$ in \eqref{eqn:fidelitymain}. This formula means that the susceptibility transition happens for the region and its complement simultaneously. We can compare these analytic results to numerical calculations for a specific choice of $\Delta$'s and $J$.

\subsection{Numerical results for $\Delta_{\phi} = 1/2,\ \Delta_O = 1/8,\ J=3/2$}
Computing the susceptibility in \eqref{eqn:fidelitymain} for general $\Delta_{\phi}$ and $1<J<3/2$ is quite difficult, even numerically. The reason is that in order to compute the susceptibility, one needs to compute $\mF(t_L,t_R;s)$ as given in \eqref{mainresult}. This requires computing $F_J(s)$ in \eqref{eqn:FullFJ}, which as far as we could tell cannot be done analytically for general $J$ and $\Delta_O$. Thus, in order to compute the susceptibility numerically one would need to compute three coupled integrals: one for the $s$ integrals in \eqref{eqn:fidelitymain}, one for the $p_+$ integral in \eqref{mainresult} and one for the $s'$ integral in \eqref{eqn:FullFJ}.

Thankfully, there are special values of $\Delta_O, \Delta_{\phi}$ and $J$ such that $F_J(s)$ can be computed analytically as well as the $p_+$ integral in \eqref{mainresult}. This leaves only the $s$-integrals in \eqref{eqn:fidelitymain} that need to be done numerically.

\begin{figure}
 \centering
 \includegraphics[width = \textwidth]{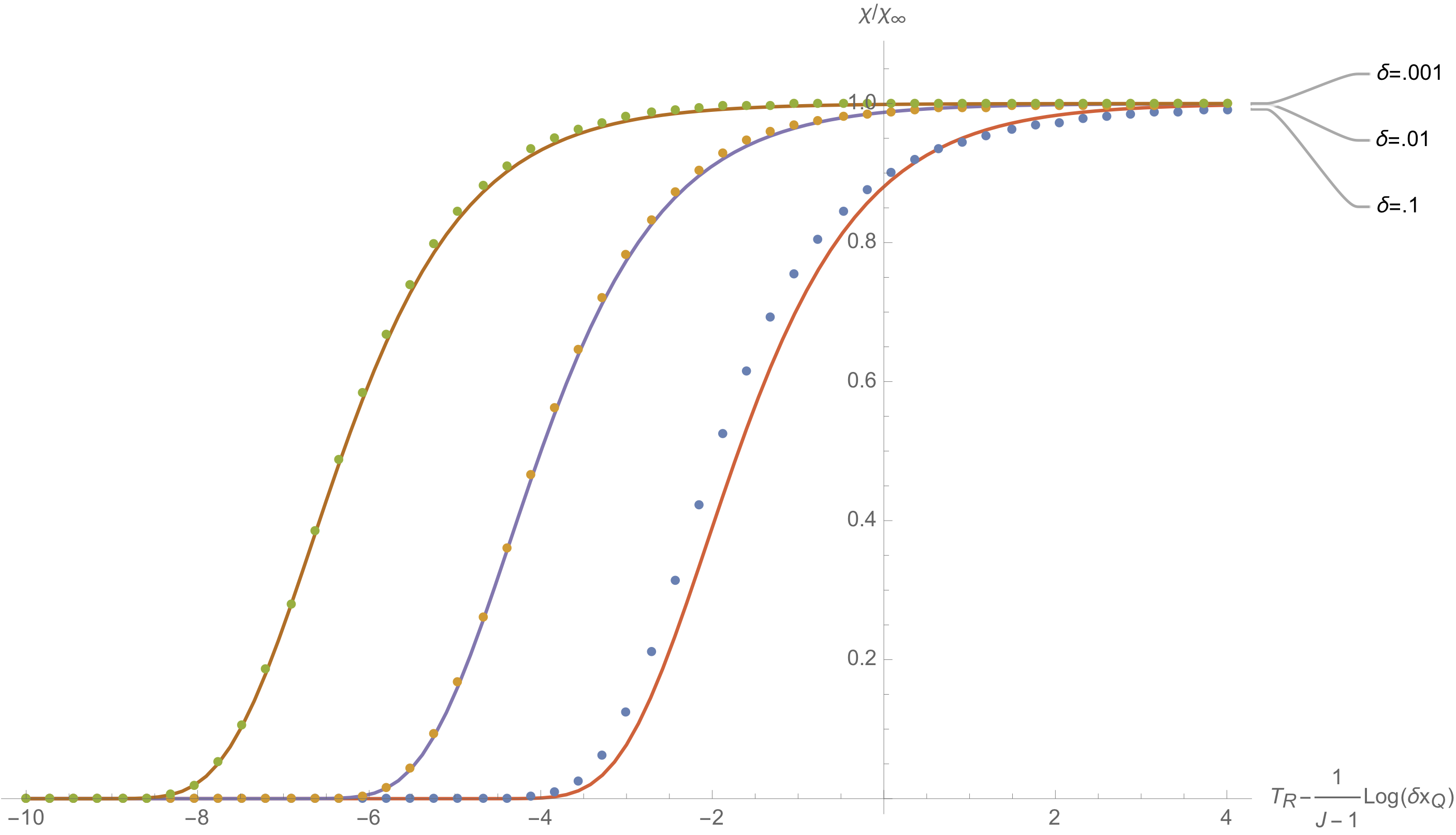}
 \caption{We plot the fidelity susceptibility normalized by its value at $T_R = \infty$ for $J=3/2$, $\Delta_{\phi} =1/2$ and $\Delta_O = 1/8$. We have shifted the time axis so that the origin corresponds to $e^{(J-1)T_R} = \delta x_Q^{(J)}$ where $\delta x_Q^{(J)}$ is defined in \eqref{eqn:xshift} as the crossover time for the full modular energy. The dots are numerically generated values and the lines are given by equation \eqref{eqn:saddlefinal} for the various $\delta$'s. We see fairly good agreement with the answer found via saddle point methods.}
 \label{fig:mainresultstringy}
\end{figure}

The values we pick are $\Delta_{\phi} = 1/2$, $\Delta_O = 1/8$ and $J=3/2$. In this case, we find
\begin{align}
F_{3/2}(s) = -\frac{e^{-i\pi 1/4}}{2\pi 2^{2\Delta_O +2}}G_N \frac{(\dl)^{2\Delta_O}}{r_I^{2\Delta_O + 1/2}}C(\Delta_O,J) H_{3/2}(s)
\end{align}
where
\begin{align}
H_{3/2}(s) =\frac{4 \pi  e^{s/2} \left(\sinh \left(\frac{s}{4}\right)+\sinh \left(\frac{s}{2}\right)+i \cosh
   \left(\frac{s}{4}\right)-i\right)}{\left(e^{s/4}+i\right) \left(e^{s/4}+1\right)}.
\end{align}
The $p_+$ integrals can then be done analytically in terms of error functions. The results are plotted for various $\delta$ in Figure \ref{fig:mainresultstringy}. We see good numerical agreement with the saddle point results in the previous subsection.

\section{Possible Bulk Interpretations}\label{sec:bulkinterp}

In this work we have computed two measures of when an excitation that has been dropped into a black hole leaves the entanglement wedge of some boundary region. We now briefly make some comments about possible bulk interpretations of these results.\footnote{We thank Douglas Stanford for first suggesting the interpretation in terms of longitudinal string spreading.} 

\subsection{A stringy QES?}

One might be tempted to associate our results on the stringy fidelity susceptibility to a change in the quantum extremal surface
itself. In particular it is natural to expect the extremal surface to become fuzzy on the string scale and indeed the transition curves we find are smoothed out on this scale (see Figure \ref{fig:mainresultstringy} and \eqref{eqn:saddlefinal}). Since the effects we are discussing are controlled by $\ell_s$, a hypothesis is that these stringy effects could be explained by including higher curvature corrections in the quantum extremal surface prescription. Such corrections were fully detailed in \cite{Dong:2013qoa}. Higher curvature terms in the quantum extremal surface equation will only produce changes to the position $\delta x_Q^+ \sim e^{T^*_R}$ which are analytic in $\ell_s$, which is not the effect we are looking for.
On the other hand we expect the higher derivative expansion to break down on stringy scales and so the Dong corrections might not be a good comparison.

There is however a more serious problem with this interpretation, which is related to the shift in the turnover time relative to a maximally chaotic theory. The turnover time for $J<2$ depends on details of the string that was thrown in. The dimension (mass) and the smearing scale of the probe both affect the turnover time, as seen in Sections \ref{sec:4} and \ref{sec:6}. This suggests that a simple change to the quantum extremal surface prescription, which one might have expected is probe independent, won't completely capture the effect we have seen. 

Nevertheless, we can think of our calculations as providing a stringy notion of the quantum extremal surface in the following sense. A boundary observer can ascertain the average location of the quantum extremal surface in the bulk from the turnover time in the fidelity. In the gravitational case, this would correspond to the true quantum extremal surface in the bulk obtained from the standard prescription. When stringy corrections are included, the boundary observer can still use the fidelity to define the location of a ``stringy'' quantum extremal surface. The location of this surface would then depend on the boundary probe. This would be an operational definition of a stringy quantum extremal surface, albeit one that goes beyond our usual understanding of an entanglement wedge. Since this is not entirely satisfying, we turn now to a different interpretation of these results.

\subsection{Longitudinal string spreading?}
Another possibility is that we are seeing an effect which is inherent to the stringy probe itself and not to the quantum extremal surface. If we assume that the quantum extremal surface position does not change appreciably (i.e. not more than order $\ell_s$ in null coordinate distance), then we can use our results to try to estimate the magnitude of string spreading.

As a first guess at the amount of longitudinal string spreading, we can imagine finding a region $LI$ such that the string is just barely, yet \emph{definitely}, in the entanglement wedge of that region. To find such a region, we can throw the string in at some fixed time $e^{T_R}$ and then tune the size of the interval, $\dl$, and its distance from the bifurcation surface, $r_I$, such that $e^{T_R}$ is right at the crossover time for the full modular energy, which as we saw in the previous section is approximately the turnover time for the fidelity susceptibility. Namely, we want to find $\dl^*, r_I^*$ that solve the equation
\begin{align}\label{eqn:Jextremalsurface}
    e^{T_R(J-1)} = \delta^{2-J} \delta x_Q^{(J)}(\dl^*,r_I^*)
\end{align}
with $\delta x_Q^{(J)}$ given in equation \eqref{eqn:dxq}. Up to order one $\Delta_O$-dependent factors, we know that $\delta x_Q^{(J)} \sim (\dl)^{2\Delta_O}/r_I^{2\Delta_O + J-1}$. Of course, there are an infinite family of $\delta \ell^*, r^*_I$ that satisfy equation \eqref{eqn:Jextremalsurface} for fixed $T_R, \delta$. This amounts to a choice of what combination of $r_I$ and $\delta \ell_I$ we hold fixed. One natural choice is to fix $r_I$ at some $J$-independent value but tune $\delta \ell$ such that \eqref{eqn:Jextremalsurface} is satisfied. Making this choice gives an estimate on the size of the string that is just the difference between $e^{T_R}$ and the quantum extremal surface position for the region with $\dl^*$ that solves \eqref{eqn:Jextremalsurface}. We get the estimate
\begin{align}
    \delta x_{\text{string}} \approx \delta x_Q^{(2)}(\dl_*) - e^{T_R} \approx e^{T_R}\left( \frac{\alpha(\Delta_O,J)}{(r_I e^{T_R}\delta)^{2-J}} - 1\right).
\end{align}
with $\alpha(\Delta_O,J)$ some order unity coefficient which goes to $1$ as $J \to 2$. In the small $2-J$ limit, we get the logarithmic dependence 
\begin{align}\label{eqn:naiveestimate}
    \delta x_{\text{string}} \approx -(2-J) e^{T_R} \log (r_I e^{T_R}\delta) +... ,
\end{align}
where we have neglected terms which don't grow as $T_R \to -\infty$. Note, however, that the estimate of $\delta x_{\text{string}}$ depends on which solution $\delta \ell^*, r_I^*$ of \eqref{eqn:Jextremalsurface} one chooses.

This formula suggests that the size of the string is proportional to $(2-J)$. As discussed in the next subsection, $(2-J)$ is proportional to $\ell_s^2/\ell_{\text{AdS}}^2$.  Remembering that we have set $\ell_{\text{AdS}}$ to one, we have that $(2-J) \sim \ell_s^2/\ell_{\text{AdS}}$ which is much smaller than the naive estimate for the string length of $\ell_s$. For this reason it is more natural to associate $\delta x_{\text{string}}$ in \eqref{eqn:naiveestimate} with $\braket{\delta x^2}/x^+(T_R)$, where $x^+(T_R) = e^{T_R}\ell_{\text{AdS}}$ is the boundary position of the infalling string in Kruskal coordinates and $\braket{\delta x^2}$ is the squared deviation of the string. If we make this identification, then interestingly our formula looks similar to the estimate computed in \cite{Larsen_1999}, who found a squared deviation of the form
\begin{align}
    \braket{(\delta x^+)^2} \approx \ell_s e^{T_R} \log (\frac{E}{\epsilon} e^{-T_R}),
\end{align}
where $E$ is the energy of the string and $\epsilon$ is related to the time resolution of the detector measuring the string. This suggests that we further make the identifications 
\begin{align}
    E \sim 1/\delta, \ \ \ r_I \sim \epsilon.
\end{align} The first identification is reasonable: the boost energy of the string is just $1/\delta$. The second identification is more non-trivial and, if correct, tells us that our use of the quantum extremal surface as a detector comes with an inherent cut-off scale set by $r_I$.  Note that as $T_R \to -\infty$ length contraction wins out as the growth in \eqref{eqn:naiveestimate} is only logarithmic. This is different from the string spreading effect predicted for strings falling across a Rindler horizon in \cite{Susskind:1994vn, Susskind_1993}.

\begin{figure}
 \centering
 \includegraphics[width = .8\textwidth]{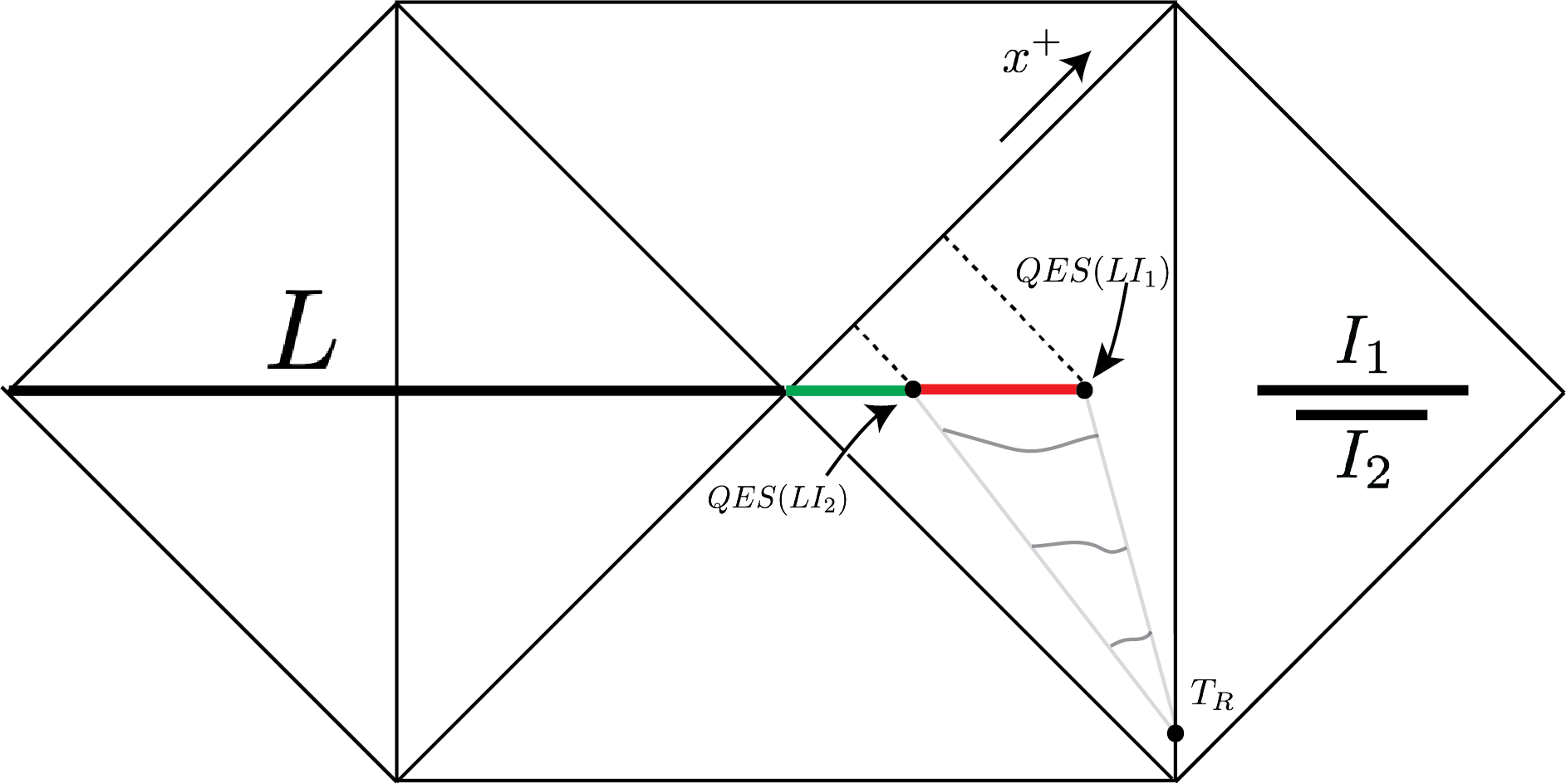}
  \caption{We can use our calculations of the fidelity (susceptibility) to infer the longitudinal extent of the string. For a fixed time $T_R$ at which we throw in the probe, we can find two different regions $LI_1$ and $LI_2$ for which the object is definitely in neither the entanglement wedge of $\overline{LI_1}$ nor the entanglement wedge of $LI_2$. This is illustrated here.}
  \label{fig:tworegions}
\end{figure}

From our saddle point calculations of the fidelity in Section \ref{sec:6}, we have more information about the size of the string. In the estimate in \eqref{eqn:naiveestimate}, we have assumed that the infalling string inserted at Kruskal coordinate $e^{T_R}$ is supported all the way from $\delta x_Q^{(2)}(\dl_*)$ down to $e^{T_R}$. This might not be the case. To examine this, we can consider two intervals, denoted $I_1$ and $I_2$, in the right bath centered around the same point. Both intervals will be a Kruskal-coordinate distance $r_I$ from the bifurcation surface. We can tune the size of $\dl_{1,2}/r_I$ so that the string is definitely in $LI_1$ and also definitely in $\overline{LI_2}$ (i.e. reconstructable from $LI_1$ and $\overline{LI_2}$). Quantitatively, we want to find the two intervals with ratio $\dl_1/\dl_2$ closest to one such that the fidelity susceptibility is very small for both $\overline{LI_1}$ and $LI_2$. The $\delta x^+$ size of the string is then estimated by computing (see Figure \ref{fig:tworegions}) 
\begin{align}
\delta x_{\text{string}} \sim \delta x_Q^+(\dl_1) - \delta x_Q^+(\dl_2).
\end{align}
where $\delta x_Q^+$ is given in equation \eqref{eqn:QESposition}.

As we saw in Section \ref{sec:6}, the fidelity susceptibility at very negative $T_R$ can be evaluated via saddle point. For a probe operator of dimension $\Delta_{\phi} = 1/2$, we found that the fidelity susceptibility for region $\overline{LI_1}=r_1$ is 
\begin{align}\label{eqn:fssaddle}
\chi(r_1)\approx 2\exp \left( -|F_J'(0)|^{\frac{1}{2-J}} e^{\frac{-T_R(J-1)}{2-J}}2^{3-2J}\delta\right) + \mathcal{O}(\delta).
\end{align}
The dependence on $\dl/r_I$ in the turnover time is entirely contained in $|F_J'(0)| \sim \left(\frac{\dl}{r_I}\right)^{2\Delta_O} \frac{1}{r_I^{J-1}}$.

For the complementary region, we found in \eqref{eqn:sumrule} that 
\begin{align}
\chi(LI_2)\approx 2\left(1-\exp \left( -|F_J'(0)|^{\frac{1}{2-J}}  e^{\frac{-T_R(J-1)}{2-J}}2^{3-2J}\delta\right)\right) + \mathcal{O}(\delta).
\end{align}
The turnover time, $T_R^*$, is defined to be when the argument of the exponential is one. We can quantitatively identify when the expression in \eqref{eqn:fssaddle} begins to dip and when it hits roughly zero as when the function's second derivatives in $T_R$ peak. These occur at $T_R \approx T_R^*(\dl) \pm \frac{2-J}{J-1}\delta T_{\pm}$, respectively, where $\delta T_{\pm} = \pm \log(\frac{3\pm \sqrt{5}}{2})$.

Thus, we want to find $\dl_1$ and $\dl_2$ such that
\begin{align}\label{eqn:timeregion}
T_R = T_R^*(\dl_1)-\frac{2-J}{J-1}\delta T_- = T_R^*(\dl_2) + \frac{2-J}{J-1}\delta T_+.
\end{align}
Using these equations and the formulae for $\delta \tilde{x}_Q^{(J)}$ in equation \eqref{eqn:dxq}, we find 
\begin{align}
   \Delta_O \log(\dl_1/\dl_2) \sim 2-J,
\end{align}
where we have dropped order one coefficients. Plugging this into the formulae for the position of the quantum extremal surface, we find that the null coordinate distance between the two quantum extremal surfaces scales as 
\begin{align}
    \delta x_{\text{string}}^+ \sim \frac{\dl_1^{2\Delta_O}}{r_I^{2\Delta_O + 1}}\left(1-e^{J-2}\right) \sim \frac{e^{T_R}}{(r_I e^{T_R}\delta)^{2-J}}(1-e^{J-2}) \label{stringspread}
\end{align}
where in the second approximate equality we used \eqref{eqn:timeregion} to exchange $\dl_1$ for $e^{T_R},\ \delta$ and $r_I$. In this case, for small $2-J$, the logarithmic dependence on $\delta, e^{T_R}$ appears only at order $(2-J)^2$. Note, however, that to use the saddle point methods of Section \ref{sec:6}, we needed to scale $\delta \ll 2-J$ (see the discussion around \eqref{eqn:saddlefinal}). In the limit where we hold $2-J$ fixed and make $\delta \ll 1$, then we find agreement between \eqref{stringspread} and \eqref{eqn:naiveestimate}. If we choose to scale $\delta \sim (2-J)^2 \to 0$, however, we find a discrepancy between the two. If indeed it is possible to interpret \eqref{stringspread} in terms of string spreading, then it suggests that the string actually propagates on a time-like trajectory. This would be a novel aspect of stringy physics in the bulk which (to our knowledge) has not been discovered before. It would be interesting to investigate this possibility in more detail and to understand better the discrepancy between \eqref{eqn:naiveestimate} and \eqref{stringspread}.

We should also mention one important point regarding the interpretation in terms of string spreading. We could have considered a different setup: consider throwing a probe string into vacuum AdS. We can ask about whether the string is reconstructable from the AdS-Rindler wedge of some single-component, spherical boundary region, $A$. We can go through the same calculations for the fidelity that we have considered here and we would end up with \eqref{eqn:fidelitymain} but with $r$ given by the boundary region $A$. For a boundary conformal field theory, the modular flow for $A$ is just a conformal boost. Thus, the fidelity is fixed entirely by conformal invariance up to the dimension of the operator. All the stringy effects would be in the dependence on the string length of the mass of the bulk particle. The essential qualitative aspects of the effects we have described would be absent. In other words, the dependence of the turnover time $T_R^*$ on the string length described in the previous sections is due to interactions between the string and the quantum extremal surface. 

\subsection{Comparison with previous work}
If our interpretation of equations \eqref{stringspread} and \eqref{eqn:naiveestimate} in terms of longitudinal string spreading is correct, then our results naively seem quite different from those found in previous works on this subject \cite{Dodelson:2017ub, Dodelson:2017wv, Dodelson:2015un, Susskind:1994vn}. In order to make a precise comparison with these works, however, one needs to be careful about what is playing the role of the detector in our setup.

A first guess would be that the quantum extremal surface itself is playing the role of the detector. Indeed, one might imagine computing the modular flowed correlation functions in \eqref{eqn:fidelitymain} via a replica trick similar to that discussed in \cite{Faulkner:2018vl}. The correlators can then be interpreted as a scattering experiment where the probe string is colliding with excitations from the twist operator sitting at the quantum extremal surface.

This is somewhat reminiscent of the calculations in \cite{Dodelson:2017wv}, where a space-time varying string coupling acts as a marker for the location of stringy interactions. Here we are similarly using properties of the bulk geometry to mark the location of the string. Interestingly, for very high detector resolution, the magnitude of string spreading in \cite{Dodelson:2017wv} is also logarithmic in the cut-off frequency. It would be nice to explore in future work whether this logarithmic dependence is analogous to the logarithms in \eqref{stringspread} and \eqref{eqn:naiveestimate}. One point, however, is that the results in \cite{Dodelson:2017wv} are exact in $\alpha'$ whereas the logarithm in \eqref{eqn:naiveestimate} appears only to leading order in $\alpha'/\ell^2_{\text{AdS}}$.

It should also be pointed out that we have not been careful to track the evolution of the string state as it falls toward the horizon. The calculations in \cite{Dodelson:2017ub, Dodelson:2017wv, Dodelson:2015un, Susskind:1994vn} work in the string vacuum. It is possible that we are effectively computing the magnitude of string spreading in a different string state. For these reasons, we believe that our calculations are not in obvious contradiction with previous work.\footnote{The points in this subsection are due to a discussion with Eva Silverstein.}

\section{Discussion}\label{sec:discussion}
We end with some comments on higher dimensions and the role of light ray operator expectation values in this work. We also discuss possible future directions.
\subsection{Higher Dimensions}
We have worked mostly in two dimensions in this paper since the formulae are more tractable. We should mention what happens in higher dimensions. Since in retrospect the full modular energy was a good indicator of when the particle has left the entanglement wedge of $r$, we can just focus on that quantity for now.

The setup we consider in higher dimensions is the one laid out in Section \ref{sec:2} (see Figure \ref{fig:regions3d}), where $L$ is a Rindler wedge on the boundary and $I$ is a sphere separated from the Rindler wedge by a distance $r_I$ and with radius $\dl/2$. In that case, we can run through all the steps for the full modular energy that we ran through in Section \ref{sec:4}, except with the wavefunctions and scattering phase modified to their higher dimensional form. For simplicity we can work in $\text{AdS}_3$/Rindler. Using the formulae in \cite{Shenker:2014tu}, we find that 
\begin{align}\label{eqn:higherdfullmod}
-\braket{\log \Delta_r}_{\psi} =& \frac{2\pi \Delta_{\phi}}{ \sin(\delta)}\cos(\delta) \nonumber \\
& -2^{2\Delta_{\phi}} \int_{-\infty}^{\infty} \frac{dk}{2\pi} \frac{e^{ikb}}{k^2 +1 } \delta x_Q(J(k),b) e^{-(J(k)-1)T_R} (\sin(\delta))^{1-J(k)}
\end{align}
where now the Pomeron spin depends on the transverse momentum $k$ as \cite{Shenker:2014tu}
\begin{align}
J(k) = 2 - \frac{\ell_s^2(k^2 + 1)}{2\ell_{\text{AdS}}^2} + \mathcal{O}(\ell_s^3).
\end{align}
The shift in the quantum extremal surface $\delta x_Q$ now also depends on the transverse position, $b$, of the probe operator, $\phi$ as 
\begin{align}
\delta x_Q(J(k),b) = -2K(\dl)^{2\Delta_O}G_N \ell_{\text{AdS}} (\ell_s^2/4)^{2-J(k)} \int dp_+ \psi^{\phi}_L(p_+,b) \psi^{\phi}_R(p_+,b) (-p_+)^{J(k)-1} \delta \tilde{x}_Q^{(J(k))}
\end{align}
and
\begin{align}
\delta \tilde{x}_Q^{(J(k))} =  \frac{\sqrt{\pi}\Gamma (2 \Delta_O +1) \Gamma (J(k)+2 \Delta_O )}{r_I^{2\Delta_O + J(k)-1}2^{2\Delta_O+1}\Gamma(2\Delta_O + \frac{J(k)}{2} + \frac{1}{2}) \Gamma(2\Delta_O + \frac{J(k)}{2} + 1)}\int dq_- \psi^O_R(q_-) \psi^O_L(q_-)(i q_-)^{J(k)-1}
\end{align}
where the momentum wavefunctions can be found in \cite{Shenker:2014tu}. The region $I$ is centered around the origin in transverse coordinates and so small $b$ corresponds to the probe being boosted directly into the past of the small region $I$. The important point for us will be that for small $b$ there is a saddle in the $k$ integral in \eqref{eqn:higherdfullmod} at small imaginary $k \sim ib$. We see that we effectively get the same behavior as in \eqref{eqn:fullmodmainresult} with $J = 2 - \ell_s^2/2\ell^2_{\text{AdS}} + \mathcal{O}(\ell_s^3)$. We expect to see the same type of behavior for the fidelity that we saw in this work in higher dimensions as well.

\subsection{Relationship between modular flowed correlator and light ray operators}
In the work of \cite{Levine:2020wz}, the computation of the correlator in \eqref{mainquant} was interpreted as a scattering process between the $\phi$ particle and a shock of size
\begin{align}
    \braket{T_{--}(x^-)} \sim \delta (x^-) H'(L:I)(e^s-1),
\end{align}
where $H(L:I) = S(L) + S(I) - S(LI)$ is the mutual information between $L$ and $I$ and $H'(LI)$ is the derivative of this quantity with respect to moving $L$'s endpoint in the outward null direction. As seen in \eqref{eqn:F2final} and discussed more generally in \cite{Levine:2020wz}, this shock has just the right value to shift the particle from the edge of the entanglement wedge of $LI$ into the causal wedge of $LI$ in the limit that $s \to -\infty$. This shock appears because, in computing \eqref{mainquant}, we are effectively acting on the vacuum with the operator $\rho_{LI}^{is} \rho_L^{-is} \rho_I^{-is}$, which the authors in \cite{Levine:2020wz} noted is the \emph{Connes cocycle} between the vacuum and a \emph{split} vacuum on $LI$. In the state $\ket{\Omega_s} = u_s(\Omega|S_{\Omega})\ket{\Omega}$ where $u_s(\Omega|S_{\Omega}) \equiv \rho_L^{-is} \rho_I^{-is} \rho_{LI}^{is}$, \cite{Levine:2020wz} found that  
\begin{align}
    \int_{0^-}^{\infty} \braket{T_{--}(x^-)}_{\Omega_s} \sim H'(L:I) (e^s-1)
\end{align}
where $x^-=0$ is the entangling surface for $L$ as in Fig. \ref{fig:ads2regions}.

For $J<2$, we lose the interpretation of \eqref{mainquant} in terms of scattering off a gravitational shockwave. Instead, we find that the string is scattering off a ``stringy'' shock, such as those discussed in \cite{Kologlu:2019uu}. Thus it is natural to guess that the function $F_J(s)$ defined in \eqref{eqn:FullFJ} represents the expectation value of a continuous spin null energy in the state excited by the cocycle, $\rho_{LI}^{is} \rho_L^{-is} \rho_I^{-is}$. Roughly speaking, we would like 
\begin{align}
    F_{J_*}(s) \sim \int_{0^-}^{\infty}dx^- \left. \braket{\mathcal{J}_J(x^-)}_{\Omega_s}\right\vert_{J\to J_*}
\end{align}
where $\mathcal{J}_J$ is the conformal primary operator of even spin $J$ on the same Regge trajectory as the stress tensor. The expression on the right hand side should then be thought of as an analytic continuation in spin $J$ from even spins to $1\leq J_*<2$.

We can understand this expression by expanding $u_s$ in the small interval limit. Then formally we have
\begin{align}
u_s = 1+ \delta^{(1)} u_s + \delta^{(2)} u_s + ....
\end{align}

Expanding for even $J$, we get
\begin{align}
    \Braket{u_s^{\dagger} \mathbb{O}_J^{\text{half}} u_s} = &\Braket{\left[\mathbb{O}_J^{\text{half}},\delta^{(1)}u_s\right]}+ \Braket{\delta^{(2)}u_s^{\dagger}\mathbb{O}_J^{\text{half}}} - \Braket{\delta^{(1)}u_s\mathbb{O}_J^{\text{half}} \delta^{(1)}u_s} + \Braket{\mathbb{O}_J^{\text{half}} \delta^{(2)}u_s},
\end{align}
where we have defined $\mathbb{O}_J^{\text{half}} = \int_0^{\infty} dx^- \mathcal{J}_J(x^-)$. Using the fact that $\delta^{(2)} u_s^{\dagger} + \delta^{(2)} u_s = \left(\delta^{(1)} u_s\right)^2$ which follows from expanding the equation $u_s^{\dagger} u_s =1$ to second order, we can rewrite this as
\begin{align}\label{eqn:expvaluecommutator}
    \Braket{u_s^{\dagger} \mathbb{O}_J^{\text{half}} u_s} = &\Braket{\left[\mathbb{O}_J^{\text{half}},\delta^{(1)}u_s\right]}- \Braket{\delta^{(1)}u_s \left[\mathbb{O}_J^{\text{half}}, \delta^{(1)}u_s\right]} + \Braket{\left[\mathbb{O}_J^{\text{half}} ,\delta^{(2)}u_s\right]}.
\end{align}

If we write $\delta^{(1)} u_s$ schematically as $O_L O_I$, where we have suppressed the integrals, and $\delta^{(2)} u_s \sim O_L O_L O_I O_I$, then we can see that the final term vanishes, since to leading order in the small interval expansion it just becomes
\begin{align}
\Braket{\left[\mathbb{O}_J^{\text{half}} ,\delta^{(2)}u_s\right]} \approx \braket{O_I O_I} \braket{[\mathbb{O}_J^{\text{half}},O_LO_L]} = \braket{O_I O_I}\braket{[\mathbb{O}_J,O_LO_L]}=0.
\end{align}

Thus, we are just left with the two first terms on the right hand side of \eqref{eqn:expvaluecommutator}. It is not hard to see that these two terms correspond to the two non-trivial terms in the right hand side of \eqref{BCH}. Thus, we get the pleasingly simple interpretation of $F_J(s)$ as the expectation value of the continuous spin null ``energy'' in the Connes cocycle flowed state $u_s(\Omega|S_{\Omega})\ket{\Omega}$.

\subsection{Conclusion}
We have examined stringy corrections to the full modular energy and fidelity for a region $L \cup I$ which includes the whole left side of a black hole together with a small subregion of the right black hole in a state with a probe string. We have found that in the limit of $\ell_s \to 0$, there is a sharp transition between when the particle is inside or outside the entanglement wedge of $LI$. In string theory, this transition gets smoothed out. We suggest that this can be interpreted in terms of longitudinal string spreading. There are many open directions. For example, it would be interesting to try to use our calculations to predict the existence of quantum extremal surfaces in various models where the bulk dual is not entirely under control (such as in SYK). It would also be interesting to examine whether there is a connection between the information theoretic quantities we have discussed and notions of scrambling involving operator size \cite{Qi:2018wg, Haehl:2021tx, Haehl:2021ta}. Both are sensitive to the bulk momentum of the probe particle and so it would be nice if there were a connection between these ideas purely within the boundary theory. We leave a more detailed discussion of these points for future work.

\section*{Acknowledgements}
We thank Ahmed Almheiri, Horacio Casini, Felix Haehl, Lampros Lamprou, Nima Lashkari, Henry Lin, Roberto Longo, Juan Maldacena, Geoff Penington, Eva Silverstein, Douglas Stanford, Alex Streicher and Ying Zhao for helpful discussions. We thank Ying Zhao for helpful comments on the draft. We especially thank Eva Silverstein for explaining previous work on longitudinal string spreading and for insightful discussions. AL acknowledges support from NSF grant PHY-1911298 and Carl P. Feinberg. V.C. is supported in part by the Berkeley Center for Theoretical Physics; by the Department of Energy, Office of Science, Office of High Energy Physics under QuantISED Award DE-SC0019380 and under contract DE-AC02-05CH11231; and by the National Science Foundation under grant PHY1820912. TF acknowledges support from the Depart of Energy, award number DE-SC0019183.

\appendix

\section{Review of modular operators}\label{app:tomitareview}
In this section we briefly summarize the relevant aspects of the Tomita-Takesaki formalism. For more details see \cite{Witten:2018wc}. Throughout this section we consider a Cauchy slice $\Sigma$ of Minkowski spacetime. Consider a subregion $\mathcal{U} \subset \Sigma $, and denote by $\mathcal{A}_{\mathcal{U}}$ the algebra of operators associated to this subregion. We denote the commutant algebra by $\mathcal{A}'_{\mathcal{U}}$. Given a state $|\psi\rangle$, the Tomita operator associated to $\mathcal{A}_{\mathcal{U}}$ is defined by 
\begin{align}\label{tomita}
S_{\psi; \mathcal{U}}\alpha |\psi\rangle  = \alpha^{\dagger}|\psi\rangle, ~ \forall \alpha \in \mathcal{A}_{\mathcal{U}}
\end{align}
Since $S_{\psi; \mathcal{U}}$ is invertible, it has a polar decomposition: 
\begin{align}\label{polar}
S_{\psi; \mathcal{U}} = J_{\psi; \mathcal{U}}\Delta^{1/2}_{\psi; \mathcal{U}}
\end{align}
where $J_{\psi; \mathcal{U}}$ and $\Delta_{\psi; \mathcal{U}}$ are the modular conjugation operator and modular operator, respectively. In the main text we will often take $|\psi\rangle = |\Omega\rangle$ and $\mathcal{U}$ to be the region $LI$ defined in Section \ref{sec:2}. 

The modular conjugation operator is anti-unitary, and also satisfies $J_{\psi; \mathcal{U}}f(\Delta_{\psi; \mathcal{U}})=\overline{f}(\Delta^{-1}_{\psi; \mathcal{U}}) J_{\psi; \mathcal{U}}$ for any function $f$. Using this, we can combine \eqref{tomita} and \eqref{polar} to get the relation 
\begin{align}\label{tomitaidentity}
J_{\psi; \mathcal{U}} \alpha |\psi\rangle = \Delta^{1/2}_{\psi; \mathcal{U}}\alpha^{\dagger}|\psi\rangle 
\end{align}
which we will make use of in the main text when $\alpha$ is the state insertion $\phi$. We also note the important commutant relations  
\begin{align}\label{commutantidentity}
J'_{\psi; \mathcal{U}} &= J_{\psi; \mathcal{U}} \nonumber 
\\ \Delta'_{\psi; \mathcal{U}} &= \Delta^{-1}_{\psi; \mathcal{U}} 
\end{align}
where $\Delta'_{\psi; \mathcal{U}}$ is defined in the same way as $\Delta_{\psi; \mathcal{U}}$ but is associated to $\mathcal{A}'_{\mathcal{U}}$, and similarly for $J'_{\psi; \mathcal{U}}$. 

We end this appendix by recalling the relationship between the modular operator and density matrices. If we have a factorizable Hilbert space $H = H_{\mathcal{U}} \otimes H_{\mathcal{U}'}$, as is the case for finite dimensional Hilbert spaces, then 
\begin{align}
\Delta_{\psi; \mathcal{U}} = \rho_{\mathcal{U}}\otimes \rho^{-1}_{\mathcal{U}'}
\end{align} 

\section{Modular Hamiltonian in the Small $\dl$ Limit}\label{app:modham}
In this Appendix, compute the analytic continuation of the sums in \eqref{eqn:reptricksum}. In particular, we focus on the term with a double sum in \eqref{eqn:reptricksum}.

\subsection{Double Sum Term}
We need to continue the term
\begin{align}
\sum_{j=0}^{n-3} \sum_{k=1}^{n-2-j} c_n(2\pi k) \text{Tr}[ \rho_L^{n-1}O^{(k+j)} O^{(j)}\psi_L \rho_L\psi_L^{\dagger}]
\end{align}

As in the main text, we write both the $j$ and $k$ sums as contour integrals. We do this first for the sum over $k$ at fixed $j$. We get
\begin{align}
&\sum_{j=0}^{n-3} \sum_{k=1}^{n-2-j} c_n(2\pi k) \text{Tr}[ \rho_L^{n-1}O^{(k+j)} O^{(j)}\psi_L \rho_L\psi_L^{\dagger}] \nonumber \\
&= \sum_{j=0}^{n-3} \frac{1}{2\pi i} \left( \oint \frac{ds_k}{e^{s_k}-1} c_n(-is_k) \times \text{Tr}[\rho_L^{n-3/2-j} O(-is_k) O \rho_L^{j+1/2} \psi \rho_L \psi^{\dagger}]\right)
\end{align}

Unwrapping the $s_k$ integral, we have a contribution from a contour when $\text{Im} s_k = 2\pi i (n-2-j)$ and one when $\text{Im} s_k = 2\pi i $. These two terms together give (at fixed $j$)
\begin{align}\label{eqn:fixedj}
&\frac{1}{2\pi i} \left( \int \frac{ds_k}{e^{s_k+i\epsilon}-1} C(-is_k+\epsilon) \times \text{Tr}[\rho_L^{n-3/2-j} O(-is_k) O \rho_L^{j+1/2} \psi \rho_L \psi^{\dagger}]\right) \nonumber \\
&-\frac{1}{2\pi i} \left( \int \frac{ds_k}{e^{s_k+i\epsilon}-1} C(-is_k+2\pi (n-2)-2\pi j+ \epsilon) \times \text{Tr}[\rho_L^{1/2} O(-is_k)\rho_L^{n-2-j}O \rho_L^{j+1/2}\psi \rho_L \psi^{\dagger}]\right).
\end{align}
Now, note that for $j=n-2$, these two terms cancel off each other. Thus, we can move the upper limit on the $j$ sum to $n-2$ for free, which we choose to do. 

We now introduce an $s_j$ contour integral for the $j$ sum. We unwrap the $s_j$ contour and we get four terms in total. We focus on the two terms coming from the top contour contour of the $s_k$ integral.
\subsubsection*{Bottom contour of $s_k$ integral in \eqref{eqn:fixedj}}

Unwrapping the $s_j$ contour on the bottom contour of the $s_k$ integral, we get 
\begin{align}
&-\frac{1}{4\pi^2} \left( \int \frac{ds_j ds_k}{(e^{s_j-i\epsilon}-1)(e^{s_k+i\epsilon}-1)} C(-is_k+\epsilon) \times \text{Tr}[\rho_L^{n-3/2} O(-is_k-is_j) O(-is_j) \rho_L^{1/2} \psi \rho_L \psi^{\dagger}]\right) \nonumber \\
&+\frac{1}{4\pi^2}\left( \int \frac{ds_j ds_k}{(e^{s_j+i\epsilon}-1)(e^{s_k+i\epsilon}-1)} C(-is_k+\epsilon) \times \text{Tr}[\rho_L^{1/2} O(-is_k-is_j) O(-is_j) \rho_L^{n-3/2} \psi \rho_L \psi^{\dagger}]\right).
\end{align}
To get something that can be analytically continued to $n=1$, we shift $s_j \to s_j-i\pi+i\epsilon$ in the first term and $s_j \to s_j+i\pi-i\epsilon$ in the second term. We then take the $n\to 1$ limit and strip off leading term. We get
\begin{align}
&\frac{-(n-1)}{2\pi} \left( \int \frac{ds_j ds_k}{(e^{s_j}+1)(e^{s_k+i\epsilon}-1)} C(-is_k+\epsilon) \times \text{Tr}[[H, O(-is_k-is_j) O(-is_j)] \psi \rho_L \psi^{\dagger}]\right) \nonumber \\
& = \frac{(n-1)}{2\pi i} \left( \int \frac{ds_j ds_k}{(e^{s_j}+1)(e^{s_k+i\epsilon}-1)} C(-is_k+\epsilon) \times \left. \frac{d}{dt}\right \vert_{t=0} \text{Tr}[ O(-is_k-is_j-it) O(-is_j-it) \psi \rho_L \psi^{\dagger}]\right) 
\end{align}
Integrating by parts, we get for the first term in \eqref{eqn:fixedj} summed over $j$ becomes
\begin{align}
& \frac{(n-1)}{2\pi i} \left( \int \frac{ds_j ds_k}{4\cosh^2(s_j/2)(e^{s_k+i\epsilon}-1)} C(-is_k+\epsilon) \times \text{Tr}[ O(-is_k-is_j) O(-is_j) \psi \rho_L \psi^{\dagger}]\right) +\mathcal{O}((n-1)^2).
\end{align}

\subsubsection*{Lower contour of the $s_k$ integral in \eqref{eqn:fixedj}}
Unwrapping the second term we get
\begin{align}
&\frac{1}{4\pi^2} \left( \int \frac{ds_j ds_k}{(e^{s_j-i\epsilon}-1)(e^{s_k+i\epsilon}-1)} C_n(-is_k+2\pi (n-2)+is_j+ \epsilon) \times \text{Tr}[\rho_L^{1/2} O(-is_k)\rho_L^{n-2}O(-is_j) \rho_L^{1/2}\psi \rho_L \psi^{\dagger}]\right) \nonumber \\
& - \frac{1}{4\pi^2} \left( \int \frac{ds_jds_k}{(e^{s_j+i\epsilon}-1)(e^{s_k+i\epsilon}-1)} C_n(-is_k+is_j+ \epsilon) \times \text{Tr}[\rho_L^{1/2} O(-is_k)O(-is_j) \rho_L^{n-3/2}\psi \rho_L \psi^{\dagger}]\right)
\end{align}

Again, in order to make this something that can be continued to $n=1$, we shift around the contours. In the top line, we shift $s_j \to s_j-i\pi$ and $s_k \to s_k+i\pi$. In the bottom line we shift $s_k \to s_k +i\pi$ and $s_j \to s_j +i\pi$. We then shift the $s_k$ contour $s_k \to s_k + s_j$ and take the $n\to 1$ limit to get
\begin{align}
&\frac{(n-1)}{2\pi} \left( \int \frac{ds_j ds_k}{(e^{s_j}+1)(e^{s_k+s_j}+1)} \text{Tr}[\rho [H,O(-is_k)] O] \times \text{Tr}[O(-is_k-is_j)O(-is_j)\psi \rho_L \psi^{\dagger}]\right) \nonumber \\
&-\frac{(n-1)}{2\pi} \left( \int \frac{ds_j ds_k}{(e^{s_j}+1)(e^{s_k+s_j}+1)} \text{Tr}[\rho O(-is_k) O] \times \text{Tr}[O(-is_k-is_j)[H,O(-is_j)]\psi \rho_L \psi^{\dagger}]\right).
\end{align}
After converting $H$ commutators to time derivatives and then integrating by parts as we did above, we find that the second term in \eqref{eqn:fixedj} summed over $j$ becomes
\begin{align}
&\frac{(n-1)}{2\pi i} \left( \int \frac{ds_j ds_k}{4\cosh^2(s_j/2)(e^{s_k+s_j}+1)}  \text{Tr}[\rho O(-is_k+\epsilon) O] \times \text{Tr}[O(-is_k-is_j)O(-is_j)\psi \rho_L \psi^{\dagger}]\right).
\end{align}

\subsubsection*{Adding the top and bottom contours of the $s_k$ integral}
Putting this all together we get
\begin{align}
&\sum_{j=0}^{n-3} \sum_{k=1}^{n-2-j} c_n(2\pi k) \text{Tr}[ \rho_L^{n-1}O^{(k+j)} O^{(j)}\psi_L \rho_L\psi_L^{\dagger}]\nonumber \\
&\sim \frac{(n-1)}{2\pi i} \int \frac{ds\, ds'}{4\cosh^2(s/2)} \left(\frac{1}{e^{s'+i\epsilon}-1} + \frac{1}{e^{s'+s}+1}\right) \times \nonumber \\
&  \text{Tr}[\rho O(-is'+\epsilon) O] \times \text{Tr}[O(-is'-is)O(-is)\psi \rho_L \psi^{\dagger}] + \mO((n-1)^2).
\end{align}

\subsection{Computing descendant contributions to $\delta H_{LI}$}
In this subsection, we compute the OPE coefficients for the descendant contributions to the twist-anti-twist OPE. We will verify equation \eqref{eqn:descopecoeff}. As discussed in the main text, we just need to expand the correlator 
\begin{align}\label{eqn:expansionapp}
&\braket{\sigma_{-n} \sigma_n O^{(j)}(z) O^{(k)}(z')}  \sim \braket{\sigma_{-n} \sigma_n} \times \nonumber \\
& =  \braket{\sigma_{-n} \sigma_n} \times  (\dl)^{2\Delta_O} \sum_{a,b=0}^{\infty} (\dl)^{a+b} c^{a,b}_{j,k} \frac{(-1)^{a+b} \Gamma(2\Delta_O+1)^2}{\Gamma(2\Delta_O +1-n)\Gamma(2\Delta_O + 1-m)} \frac{1}{z^{2\Delta_O+a} (z')^{2\Delta_O+b}} + ... \nonumber \\
\end{align}
as a double power series in $\dl/z,\,\dl/z'$.

The correlation function $\braket{\sigma_{-n} \sigma_n O^{(j)}(z) O^{(k)}(z')}$ can be computed by introducing the 
\begin{align}
\tilde{x}(z) = \left(x(z)\right)^{1/n} =  \left(\frac{z-\dl/2}{z+\dl/2}\right)^{1/n}
\end{align}
which implements the insertion of the (anti-)twist operators at $z = \pm \dl/2$. We can then compute $\braket{\sigma_{-n} \sigma_n O^{(j)}(z) O^{(k)}(z')}$ by conformally transforming a vacuum two-point function of $O$'s from $\tilde{x}$ coordinates to $z$ coordinates. We get
\begin{align}\label{eq:preexpand}
\braket{\sigma_{-n} \sigma_n O^{(j)}(z') O^{(0)}(z)} = \frac{1}{n^{2\Delta_O}}\left(\frac{dx}{dz} x^{1/n-1}\right)^{\Delta_O}\left(\frac{dx'}{dz'} (x')^{1/n-1}\right)^{\Delta_O} \frac{e^{2\pi i j\Delta_O/n}}{\left((x')^{1/n}e^{2\pi i j/n}-x^{1/n} \right)^{2\Delta_O}}
\end{align}

We then expand this answer to leading order in $\dl/z'$ and to all orders in $\dl/z$, dropping terms that are proportional to $n-1$, since these will not contribute in the $n \to 1$. We find
\begin{align}
&\braket{\sigma_{-n} \sigma_n O^{(j)}(z') O^{(0)}(z)} \nonumber \\
&= \frac{(\dl)^{2\Delta_O}}{n^{2\Delta_O}(zz')^{2\Delta_O}}\frac{1}{\left(e^{\pi i j/n} - e^{-i\pi j /n}\right)^{2\Delta_O}}\sum_{b=0}^{\infty} \left( \frac{\dl}{2z}\right)^{b} \frac{(-1)^b \Gamma(2\Delta_O +1)}{b! n^b\Gamma(2\Delta_O +1-b)} \left( \frac{e^{\pi i j/n} + e^{-i\pi j/n}}{e^{\pi i j/n}- e^{-i\pi j/n}}\right)^b \nonumber \\
 &+ \text{(terms proportional to $n-1$}).
 \end{align}

Comparing with \eqref{eqn:expansionapp} above, we see that
\begin{align}
c_{j,0}^{a=0,b} = \frac{1}{\left(e^{\pi i j/n} - e^{-i\pi j /n}\right)^{2\Delta_O}}\left( \frac{e^{\pi i j/n} + e^{-i\pi j/n}}{e^{\pi i j/n}- e^{-i\pi j/n}}\right)^b \frac{1}{b! 2^b n^{b+2\Delta_O}} + \text{(terms proportional to $n-1$)}
\end{align}
which agrees precisely with \eqref{eqn:descopecoeff}.

\section{Conformal factors from embedding space}\label{app:conformalfactor}
In this Appendix, we briefly review the embedding space formalism for CFTs and then argue for equation \eqref{eqn:conformalfactor} in the main text. Conformal field theories in $(d-1)+ 1$ dimensions can be viewed as living on the null cone in $d+2$ dimensions defined by $P \cdot P = 0$ where $P \in \mathbb{R}^{d,2}$ together with the condition that rescalings $\lambda P \sim P$ are pure gauge. 

Since our calculations in the main text are mainly for $d=2$, we focus on that case, although everything we say can be upgraded to higher dimensions. There are two frames or gauge choices we will be interested in. In Lorentzian signature, the embedding space metric is 
\begin{align}
ds^2 = -dP_I^2 + dP_{II}^2 -dP_0^2.
\end{align}
Then the two choices are the flat frame
\begin{align}
P \vert_F = \left(\frac{\frac{\dl^2}{4}-x^2}{\dl}, \frac{\frac{\dl^2}{4}+x^2}{\dl}, x\right)
\end{align}
and hyperbolic
\begin{align}
P \vert_H = (1, \cosh(t), \sinh(t)).
\end{align}

The conformal transformation between these two frames maps the interval $x \in (-\dl/2,\dl/2)$ to the hyperbolic line. If we define $\Omega$ such that 
\begin{align}
\Omega P \vert_F = P \vert_H
\end{align}
we get that 
\begin{align}
\Omega = 2 (1+ \cosh(t))/\dl = (2\cosh(t/2))^2/\dl = \frac{\dl}{\dl^2/4-x^2}.
\end{align}

The two-point function of two scalar conformal primaries of dimension $\Delta_O$ in a given frame is 
\begin{align}\label{eqn:twopf}
\braket{O(P_1) O(P_2)} = \frac{1}{(P_1 \cdot P_2)^{\Delta_O}}.
\end{align}
Conformal boosts for the region $I$ corresponds to just time translation in the hyperbolic frame. If we denote the boosts by $M(t)$, then we have
\begin{align}\label{eqn:transformation}
M(t)\cdot P\vert_F = (M(t) \cdot P)\vert_F \times \frac{\Omega(t)}{\Omega(0)}
\end{align}
where this equation follows from having to map $P \vert_F$ to the hyperbolic frame, time evolve and then map back to the flat frame. The conformal boost keeps us within the flat frame up to the conformal transformation $\tilde{\Omega}(0,t)$ where
\begin{align}
\tilde{\Omega}(s,t) = \frac{\Omega (s)}{\Omega(t+s)} = \frac{\cosh^2(s/2)}{\cosh^2((s+t)/2)}.
\end{align}

The point $(M(t) \cdot P)\vert_F$ corresponds to the flow 
\begin{align}
x^-(s) &= -r + \frac{\dl}{2} \left( 1+ \frac{2(x^-+r -\dl/2)}{-r+\dl/2-x^-+e^s(x^-+r+\dl/2)}\right).
\end{align}
We can introduce the Mobius transformed coordinate
\begin{align}
\overline{x} = -r_- \frac{x + r_+}{x+r_-}
\end{align} 
which maps points $x>0$ to points $x \in (-r_+, -r_-)$ where $r_{\pm} = r_I \pm \dl/2$. One can that if $x^-(0) = \overline{re^s}$, then 
\begin{align}\label{eqn:flow}
x^-(t) = \overline{re^{t+s}}.
\end{align}
Putting equations \eqref{eqn:flow} and \eqref{eqn:transformation} into \eqref{eqn:twopf}, we find that
\begin{align}
e^{iH_I t} O_I(\overline{re^s}) e^{-iH_It} = \tilde{\Omega}^{\Delta_O}(s,t) O_I(\overline{re^{s+t}}).
\end{align}

\section{Computing the fidelity susceptibility}\label{app:susceptibility}
In this appendix, we write down the fidelity susceptibility for perturbatively differing states in terms of modular operators, using the result in \cite{May:2018ti}, which was derived in terms of density matrices. For two density matrices $\rho, \sigma$ on some region $r$, and associated to global states $\ket{\psi}$ and $\ket{\Omega}$ respectively, the fidelity is given by 
\begin{align}
    F(\rho\vert \sigma; r) = \text{Tr}\left(\left(\sigma^{1/2}\rho \sigma^{1/2}\right)^{1/2} \right)
\end{align}
For the special case of perturbative states, i.e. 
\begin{align}
\rho \approx \sigma + \lambda \delta \rho + \mathcal{O}(\lambda^2), 
\end{align}
Hijano \& May \cite{May:2018ti} computed the fidelity to leading order in small $\lambda$, which is just the fidelity susceptibility $\chi(\rho, \sigma)$. As quoted in the main text, they found that 
\begin{align}\label{eqn:susceptibilityappendix}
&\chi(\psi,\Omega;r) = \int ds \tilde{P}_{1/2,1/2}(s) \text{Tr}[\sigma^{-1} \delta \rho \sigma^{-is/2\pi} \delta \rho \sigma^{is/2\pi}]
\end{align}
where $\tilde{P}_{1/2,1/2}(s)$ is the Fourier transform of the function
\begin{align}
P_{1/2,1/2}(\omega) = \frac{1}{1+ e^{2\pi \omega}}.
\end{align}
In particular, we have
\begin{align}
&\tilde{P}_{1/2,1/2} (s) = \frac{-i}{4\pi} \frac{1}{\sinh((s-i\epsilon)/2)}.
\end{align}

Our goal is now to argue for equation \eqref{eqn:fidelityoperatorinsertion} in the main text. We consider states $\rho$ such that the perturbation $\delta \rho$ is given by the insertion of an operator $O$ in the Euclidean time plane. Namely,
\begin{align}
\delta \rho = \sigma^{1-\alpha} O^{\dagger} \sigma^{\alpha} + \sigma^{\alpha} O \sigma^{1-\alpha}
\end{align}
where $\alpha <1/2$.

Inserting for $\delta \rho$ into \eqref{eqn:susceptibilityappendix}, we have two terms
\begin{align}
&\text{Tr}[\sigma^{-1} \delta \rho \sigma^{-is/2\pi} \delta \rho \sigma^{is/2\pi}] \nonumber \\
& =\text{Tr}[\sigma_r (O_r(\alpha))^{\dagger} \sigma^{is/2\pi} \sigma^{-1}\delta \rho \sigma^{-is/2\pi}] + \lambda^2 \text{Tr}[\sigma \sigma^{is/2\pi} \sigma^{-1} \delta \rho \sigma^{-is/2\pi}O(\alpha)]
\end{align}
with $O_r(\alpha) = \sigma^{\alpha} O_r \sigma^{-\alpha}$ and where we have used cyclicity of the trace.

These are just correlators in the state $\sigma$ and so we have 
\begin{align}
&\text{Tr}[\sigma^{-1} \delta \rho \sigma^{-is/2\pi} \delta \rho \sigma^{is/2\pi}] 
& =\lambda^2 \left( \braket{ (O_r(\alpha))^{\dagger} \sigma^{is/2\pi} \sigma^{-1}\delta \rho \sigma^{-is/2\pi}} + \braket{\sigma^{is/2\pi} \sigma^{-1} \delta \rho \sigma^{-is/2\pi}O_r(\alpha)}\right)
\end{align}
where these are correlators in the purification of the state $\sigma$ onto the full system.

Now,
\begin{align}
&\sigma^{is/2\pi} \sigma^{-1} \delta \rho \sigma^{-is/2\pi} \ket{\Omega} = \Delta^{is/2\pi} (\Delta^{\alpha-1} O + \Delta^{-\alpha} O^{\dagger})\ket{\Omega}
\end{align}
and when we plug this back in, we can do the $s$-integral to get
\begin{align}
\frac{1}{1+\Delta} (1+\Delta^{1/2}J) \ket{\delta \psi}
\end{align}
with
\begin{align}
\ket{\delta_{\lambda}\psi} = \Delta_r^{\alpha} O_r\ket{\Omega}.
\end{align}

Furthermore,
\begin{align}
\bra{\Omega} \sigma^{is/2\pi} \sigma^{-1} \delta \rho \sigma^{-is/2\pi} = \bra{\Omega} (O\Delta^{1-\alpha} + O^{\dagger} \Delta^{\alpha}) \Delta^{-is/2\pi}.
\end{align}
We can rewrite this expression in terms of the Tomita operator $S$, which implements Hermitian conjugation $S O \ket{\Omega} = O^{\dagger} \ket{\Omega}$ and also we can remember $J \Delta^{\alpha}J = \Delta^{-\alpha}$ for $\alpha$ real.
Thus, we get
\begin{align}\label{eqn:susceptibilityoverlap}
\chi(\psi,\Omega;r) =2\text{Re} \braket{\delta \psi| \tilde{P}_r |\delta \psi}
\end{align}
where
\begin{align}
\tilde{P}_r = \frac{1}{1+\Delta} (1+ \Delta^{1/2} J).
\end{align}

We would like to rewrite this in terms of modular flow, so we can go back to the spectral representation, where \eqref{eqn:susceptibilityoverlap} can be written as
\begin{align}
&\chi(\psi,\Omega;r)= \text{Re} \frac{-1}{2\pi i } \int ds \left( \frac{\braket{\delta \psi| \Delta^{is/2\pi}|\delta \psi}}{\sinh((s+i\epsilon)/2)} - i \frac{\braket{\delta \psi| \Delta^{is/2\pi}J|\delta \psi}}{\cosh(s/2)}\right),
\end{align}
which is equation \eqref{eqn:fidelityoperatorinsertion}.

\bibliographystyle{JHEP}
\bibliography{all}
  
\end{document}